\documentstyle[11pt,aaspp4]{article}

\slugcomment{Accepted for ApJ}

\lefthead{Yi et al.}
\righthead{The $Y^2$ Isochrones}

\begin{document}

\title{Towards Better Age Estimates for Stellar Populations:\\ The $Y^2$ Isochrones for Solar Mixture}

\author{Sukyoung Yi\altaffilmark{1}}
\affil{Center for Space Astrophysics, Yonsei University, Seoul 120-749, Korea and \\ California Institute of Technology, Mail Code 405-47, Pasadena, CA 91125\\ yi@srl.caltech.edu}
\author{Pierre Demarque}
\affil{Department of Astronomy, Yale University, PO Box 208101, New Haven, CT 06520-8101 \\ demarque@astro.yale.edu}
\author{Yong-Cheol Kim, Young-Wook Lee, Chang H. Ree}
\affil{Center for Space Astrophysics, Yonsei University, Seoul 120-749, Korea \\ kim@csa.yonsei.ac.kr, ywlee@csa.yonsei.ac.kr, chr@csa.yonsei.ac.kr}
\author{Thibault Lejeune}
\affil{Observatorio Astronomico, Universidade de Coimbra, Santa Clara 3040 Coimbra, Portugal \\ lejeune@mat.uc.pt}

\and

\author{Sydney Barnes}
\affil{Department of Astronomy, University of Wisconsin, 475 N Charter Street, Madison, WI 53706 \\ barnes@astro.wisc.edu}

\altaffiltext{1}{Current address: University of Oxford, Astrophysics, Keble Road, Oxford, OX1 3RH, United Kingdom} 

\begin{abstract}

We have constructed a new set of isochrones, called the $Y^2$ Isochrones, that
represent an update of the Revised Yale Isochrones (RYI), using improved
opacities and equations of state.
	Helium diffusion and
convective core overshoot have also been taken into consideration.
	This first set of isochrones is for the scaled solar mixture. A 
subsequent paper will consider the effects of $\alpha$-element enhancement,
believed to be relevant in many stellar systems. 
	Two additionally significant features of these isochrones are that
(1) the stellar models start their evolution from the pre-main sequence
birthline instead of from the zero-age main sequence, and (2) the
color transformation has been performed using both the latest table of
Lejeune et al., and the older, but now modified, Green et al. table.
	The isochrones have performed well under the tests conducted thus far.
	The reduction in the age of the Galactic globular clusters caused 
by this update in stellar models alone is approximately 15\% relative to 
RYI-based studies.
	When the suggested modification for the $\alpha$-element enhancement
is made as well, the total age reduction becomes approximately 20\%.
	When post-RGB evolutionary stages are included, we find that 
the ages of globular clusters derived from
integrated colors are consistent with the isochrone fitting ages.

\end{abstract}


\section{Introduction}

	Isochrones are defined as the locus of coeval (equal age) points on
the evolutionary tracks of stars of different masses in the
Hertzsprung-Russell Diagram (HRD).
	An older isochrone has a fainter and redder main-sequence
turn-off (MSTO), because the brighter and bluer massive stars evolve and die
earlier.
	For the past four decades, astronomers have used this basic fact in
various ways to derive the ages of star clusters and galaxies.

        The first systematic application of the isochrone method was made
to NGC\,188 MSTO by Demarque and Larson (1964), as noted by Sandage and Eggen
(1969).
	The widespread adoption of the isochrone technique has been
tremendously helpful in understanding the formation and evolution of the Milky
Way and its components, and the chronological sequence of astrophysical
processes therein (see VandenBerg, Bolte, \& Stetson 1996;
Sarajedini, Chaboyer, \& Demarque 1997).
	Today, isochrones are the most powerful means of measuring
the ages of star clusters, and thanks to numerous recent improvements
in the input physics, we may now claim to know the age of the Milky Way
better than ever before.
	This purely theoretical achievement helps setting  one of the
most important constraints on cosmology, because the age of the
universe is a distinctive product of any cosmological model.

	A second-generation isochrone technique extends beyond
the MSTO up to the tip of the red giant branch (RGB), to match more stars in
the HRD (Iben 1974; Demarque \& McClure 1977).
	 The use of the extended isochrone, now standard, confers at least
two significant advantages over the use of only the MSTO.
	Firstly, isochrone fitting is no longer in principle sensitive to 
the uncertainty in distance measurement.
	This advantage, however, has not practically been much appreciated
because the uncertainties in the stellar atmosphere models
and in the convection approximations still make it difficult to reproduce
accurately the colors of red giants.
	Secondly, it allows us to build realistic population models for
clusters and galaxies from their initial mass functions.
	This technique, called ``evolutionary population synthesis'' (EPS),
was pioneered by Tinsley (1980).

In spite of the attendant difficulties, attempts to reproduce the colors
and spectra of clusters and galaxies through the EPS technique are increasingly
successful, and especially so when empirical stellar spectral libraries are
resorted to (eg. Larson and Tinsley 1978; Gunn, Stryker, \& Tinsley 1981; 
Bruzual 1983; Pickles 1985).
	The usage of isochrones has increased steadily.
	Notably, the recent studies of Spinrad et al. (1997) and
Yi et al. (2000) have demonstrated that precision-isochrones can be used to
constrain the galaxy formation epoch directly by estimating the
ages of high-redshift galaxies in their youth.
	Such studies have demonstrated that population synthesis
studies with precision-isochrones may even differentiate between various
cosmological models.

	Isochrones need to be updated when significant improvements are made
in input physics. The Yale group (eg. Green, Demarque, \& King 1987) is one
among several involved in such activity. 
	Other notable recent updates include Schaller et al. (1992), 
Straniero, Chieffi, \& Limongi (1997), Girardi et al. (2000), and 
VandenBerg et al. (2000).
	The Yale group's prior published set of isochrones, the Revised Yale 
Isochrones (\cite{gdk87}: hereafter RYI), are out-of-date.
	A number of specialized studies have been published over the years by
the Yale group which have updated the physics of stellar models
(Guenther et al. 1992, 1996; Chaboyer \& Kim 1995; Guenther \& Demarque 1997;
Yi et al. 1997; Chaboyer et al. 1998; Heasley et al. 2000).
	We note especially the importance of substantial improvements
in the opacities and equation of state (mostly originating from the OPAL 
group: \cite{ir96}).
	Because of the need for a comprehensive set of isochrones for
research in stellar populations, we hereby present a grid of Yale 
isochrones using up-to-date input physics and parameters.

	Our new isochrone set (``the $Y^2$ Isochrones'', after the
Yonsei-Yale collaboration) covers a wide range in metallicity
and age.
	A wide $Z$-coverage in particular is useful in constructing
EPS models for complex populations, such as elliptical galaxies.
	The ages in the $Y^2$ Isochrones are computed starting from
the pre-MS birthline rather than from the ``zero-age'' MS (ZAMS).
	This allows us to build realistic young isochrones that are mainly
defined by pre-MS stellar models.
	In this respect, they should be useful in population studies of
youthful systems, such as young open clusters.

	In this paper (the $Y^2$ isochrones-solar mixture), 
all isochrones are based on a chemical composition that has been scaled to 
the solar mixture.  
	The case of $\alpha$-element enhancement
will be considered in a subsequent paper.
	This is because we believe that we do not as yet have a clear
understanding of it as a function of mass and of metallicity, particularly in
the high metallicity regime.
	If $\alpha$-element enhanced isochrones is desired, we recommend the
use of a correction formula similar to the one suggested by
Salaris et al. (1993) with this set of isochrones.  
	This correction formula has proven reliable in low metallicity 
systems, such as the globular clusters (Chaboyer et al. 1992), when compared 
to models actually constructed with opacities for
$\alpha$-element enhanced mixtures.
	Subsequent work will present $Y^2$ isochrones that
include $\alpha$-element enhancement based on more up-to-date opacities.
	Users will then be able to construct isochrones for the desired values
of the $\alpha$-element enhancement via simple interpolation.
	The corresponding post-RGB models will also be available presently.

\section{Input Physics and Parameters}

	The physics used in this study for the construction of stellar models
is summarized in Table 1 and discussed below.

\placetable{tbl-1}

\subsection{Microscopic Physics}

	 The OPAL opacities are the most widely used Rosseland mean opacities
today.
	Lately, the group has released newer tables for the Grevesse and Noels
(1993) solar mixture that include the effects of seven additional heavy
elements.
	These new calculations also include the effect of some changes in
physics (\cite{ri95}; \cite{ir96}).
	All these effects have been taken into account.
	We use the OPAL opacities for the temperature range of 
$\log T \geq 4.1$.
	For $\log T \leq 3.75$, we use the Alexander \& Ferguson opacities 
(1994).
	In the temperature region of $3.75 < \log T < 4.1$, we use a
value linearly interpolated between these two sets of tables.
	The conductive opacity is always included when 
$\log T \geq 4.2$ and  $\log \rho \geq 2 \times \log T - 13$, 
where $\rho$ is density.  We adopt the work of Hubbard \& Lampe 
(1969) for $\log \rho \leq 6.0$ and the work of Canuto (1970) 
in the relativistic regime where $\log \rho > 6.0$.  
	Deep inside the star where the OPAL opacity table fails to cover, the
electron scattering is the dominant mechanism.

	The equation of state was taken from Rogers, Swenson, \&
Iglesias (1996), i.e., OPAL EOS.
	Beyond the boundaries of the table, the standard Yale implementation
with the Debye-H\"uckel correction was used (\cite{gdkp92}; \cite{ck95}).
  	The most popular equations of state have been simple Saha solvers, 
which we used when $\log T \leq 3.7$.
	As the importance of the Coulomb interaction has been acknowledged,
the effect has been included using the Debye-H\"uckel approximation.
	Care has to be taken, however,  that the Debye-H\"uckel correction
is only applied when $\Lambda < 0.2 $ (see page 23-26 for definition of
$\Lambda$ and discussion in Rogers (1994)).
	For masses of 0.75$M_\odot$ and above, $\Lambda$ is always less
than 0.2 in the post-ZAMS stellar evolution models.
	Thus, models constructed using the OPAL equations of state
and those of the usual Saha solver together with the Debye-H\"uckel
approximation are similar (\cite{ck95}).
	Note that the treatment of Coulomb interaction in the OPAL equations
of state are valid for values of $\Lambda$ up to 3, more than an order
of magnitude better than the simple Debye-H\"uckel formula in common use.
	At very high temperature and/or pressure, where the OPAL EOS tables
do not cover, we fall back to the YREC built-in EOS, 
which follows Cox \& Giuli (1968), and is described in detail in 
Appendix A of Prather (1976).  
	Here the pressure ionization is the dominant effect, i.e. fully 
ionized medium.
	Between tables, we have interpolated linearly.

  	To include helium diffusion in the calculation, we have employed
Loeb's formula (Bahcall, \& Loeb 1990; Thoul, Bahcall, \& Loeb 1994).
	For a more detailed discussion of diffusion, see \cite{cha92} where
the work of \cite{tho94} and \cite{mp93} have been compared.
	There is an uncertainty in the diffusive coefficient of each metal
element.
 	Thus, common practice assumes that all heavy elements
diffuse with the same velocity as fully ionized iron.
	Guenther \& Demarque (1997) present the results of such computations.
	This study includes only helium diffusion.

	The energy generation routines of Bahcall \& Pinsonneault (1992)
and the cross sections listed in Bahcall (1989) have been used.
	In addition, the cross sections for the $pp$ reactions,
the $Be^{7}$-proton capture reaction, and the $hep$ reaction have been
updated (Bahcall and Pinsonneault, priv. comm.).
	These routines, kindly provided by Bahcall (priv. comm.),
also include coefficients for weak, intermediate and strong electron screening,
based on the Salpeter (1954) theory of weak screening (which is valid in most
cases in stellar evolution calculations), and its extension to
intermediate and strong screening (DeWitt et al. 1973; Graboske et al. 1973).
	Neutrino losses from photo, pair and plasma neutrinos are included
in the energy generation following the work of Itoh et al. (1989).

\subsection{Solar Calibration}

	Our stellar models have been calibrated against the sun.
These models are based on the solar mixture of Grevesse \& Noels (1993).
	Grevesse \& Noels made a major effort to determine more accurate
CNO \& Fe abundances, and now the solar Fe abundance agrees with the
meteoritic abundance better than ever before.
	Their new solar metal-to-hydrogen ratio is $(Z/X)_\odot=0.0245$,
while the previous value
from Anders \& Grevesse (1989), based on the meteoritic Fe, was 0.0267.
	The value we have finally adopted is 0.244 from the even more 
up-to-date value from Grevesse, Noels, \& Sauval (1996).

	The model that best matches the solar properties
($L_\odot$ = 3.8515E33 ergs\,$\rm s^{-1}$ and $R_\odot$ = 6.9551E10 cm) at the
generally accepted age of the sun (4.50\,Gyr) is of
$(X, Z)_{0}=(0.7149, 0.0181)$, i.e., $(Z/X)_\odot=0.0243$.
       Our solar model achieves the following agreements:
$T_{\rm eff}/T_{\rm eff \odot} = 0.999936$,
log ($L/L_{\odot}) = 0.0000000185$, and log ($R/R_{\odot}) = -0.0000006701$.
	At the solar age, the surface composition was (X, Z) =
(0.7463, 0.0181).
	The mixing length parameter of $l/H_{p} = 1.7432$ has been used to
produce this match and thus for all other models.

	Note that we have used the slightly larger value of  $L_\odot$
(3.8515E33 ergs $\rm s^{-1}$) rather than the one listed in Livingston (2000),
i.e., 3.845E33 ergs $\rm s^{-1}$.
	This 0.17\% difference in luminosity is due to the uncertainty in
the measured value of the solar irradiance.
	Guenther \& Demarque (1992) state that
``the luminosity has been determined from solar constant measurements from
space on both the Nimbus 7 and the SMM satellites (Hickey \& Alton 1983).
ERB-Nimbus measures $\rm 1371.6 \pm 0.765 W m^2$ and SMM/ACRIM measures
$\rm 1367.7 \pm 0.802 W m^2$ which yields luminosities of
3.846E33\,erg\,$\rm s^{-1}$ and 3.857E33 \,erg\,$\rm s^{-1}$, respectively.
	As we cannot expertly base a
preference, we merely take the average of the two and adopt the solar
luminosity in all our models to be $L_{\odot}$ = 3.8515E33 erg $\rm s^{-1}$.''
	The accuracy of the Nimbus 7 measurements has more recently been 
estimated to be 0.5\% (Lee et al. 1995).  
	We simply adopt the value of $L_{\odot}$ from Guenther \& Demarque 
(1992) in this study.

\subsection{Chemical Abundance and $\Delta$$Y$/$\Delta$$Z$}

	We have set the initial chemical composition to
$(Y, Z)_{0}=(0.23, 0.00)$.
	Our solar calibration (described above) suggests
the initial solar chemical composition of $(Y, Z)_{\odot}$ =
(0.267025, 0.018100).
	This indicates $\Delta$$Y$/$\Delta$$Z \approx 2.0$, i.e.,
\begin{equation}
		Y = 0.23 + 2\, Z.
\end{equation}
	This slope is the same as that from VandenBerg et al. (2000) and
only slightly smaller than that (2.25) of Girardi et al. (2000).
	Given the inevitable differences in codes and input parameters, we
regard this as a good agreement.
	Table 2 displays the chemical compositions selected for this study.

\placetable{tbl-2}

	It should be noted that $\Delta$$Y$/$\Delta$$Z$ is not a precisely
determined quantity.
	In this study, it has been determined based on two values: the initial
composition and the solar composition.
	Various techniques have indicated differing values, but
most estimates cluster between 2 and 5 (see Pagel \& Portinari 1998 for
review).
	Adoption of a slightly different value does not greatly modify
the stellar evolution when $Z$ is low.
	However, it can lead to unrealistic stellar models when
$Z$ is very large ($> 2 Z_\odot$), because helium abundance is
the prime factor in setting the pace of stellar evolution.
	In this sense, our extremely metal-rich models based on a crude
value of $\Delta$$Y$/$\Delta$$Z$ may not have the same accuracy as their
metal-poor counterparts.

\subsection{Correction for $\alpha$-Element Enhancement}

	The isochrones presented in this paper are intended for solar-type
populations that do not show signs of $\alpha$ enhancement.
	For populations with strong signs of $\alpha$ enhancement, such
as metal-poor globular clusters, we recommend the correction formula
similar to the one provided by Salaris, Chieffi, \& Straniero (1993):
\begin{equation}
	Z = Z_{0} (0.638 f_\alpha + 0.362),
\end{equation}
or, one more appropriate to the Grevesse \& Noels (1993) mixture, which is:
\begin{equation}
	Z = Z_{0} (0.694 f_\alpha + 0.306),
\end{equation}
where $f_{\alpha}$ is the chosen $\alpha$ enhancement factor, $Z_{0}$ is the
actual metallicity of the target, and $Z$ is the metallicity of the appropriate
non-$\alpha$-enhanced isochrone.
	The modification of the original Salaris et al. formula
is necessary because of the change in the mixture adopted.
	This effect is probably small, as indicated by Salaris et al. (1993). 
	But we also emphasize that such corrections are likely to be invalid 
at the higher metallicities.

Upon availability of our second set of isochrones (with $\alpha$
enhancement included), we will validate this issue and users will be able
to generate the appropriate isochrones for any reasonable $\alpha$ enhancement
factor through a simple interpolation.

\subsection{Convective Core Overshoot}

        Convective core overshoot (OS), the importance of which was first
pointed out by Shaviv \& Salpeter (1973), is the inertia-induced penetrative
motion of convective cells, reaching beyond the convective core as defined
by the classic Schwarzschild criterion.
        Stars develop convective cores if their masses are larger than
approximately 1 -- 2 $M_\odot$, typical for the MSTO stars in 1 -- 5\,Gyr-old
populations, depending on their chemical composition.
        Since the advent of the OPAL opacities, various studies have
suggested a modest amount of OS; that is, OS $\approx 0.2$ $H_{p}$ for
clusters of age 1 -- 2\,Gyr but OS $\approx 0.0$ -- 0.1 $H_{p}$ for
older clusters (4 -- 6\,Gyr), where $H_{p}$ is the pressure scale height
(\cite{s91}; Demarque, Sarajedini, \& Guo 1994; \cite{din95}; \cite{koz97}).

	In this study, we have adopted OS = 0.2 for younger isochrones
($\leq 2$\,Gyr) and OS = 0.0 for older ones ($\geq 3$\,Gyr).
	This is based on the observational studies listed above.
	It should be noted, however, that such tests of the models were
performed mainly against the approximately solar abundance populations, and
thus our adoption of the OS parameter may not be valid in chemically
different populations.

	For young isochrones with OS, we first found the critical mass
$M^{conv}_{crit}$  above which stars continue to have a
substantial convective core even after the pre-MS phase is ended, by 
inspecting the stellar models with no OS.
	The mass interval between our adjacent models being 0.1 $M_\odot$,
the actual $M^{conv}_{crit}$ is between ($M^{conv}_{crit}$ listed) and
($M^{conv}_{crit} - 0.1$).
	Then, we constructed heavier stellar models with OS included and
used them in the construction of young isochrones.
	For old isochrones, we used stellar models with no OS regardless
of mass.
        OS has many effects on stellar evolution (see Stothers 1991).
	Among the most notable are its effects on the shape of the MSTO,
on the rate of evolution near the MSTO and the subgiant phase, and 
on the ratio of total lifetimes spent in the core hydrogen burning stage (MS) 
and in the shell hydrogen burning stage (RGB).
	The impacts of such effects on the isochrone and on the integrated
spectrum have been discussed by Yi et al. (2000).

\section{Construction of Stellar Models}

\subsection{Evolutionary Tracks}

	We have used the Yale Stellar Evolution Code (YREC) to construct
these stellar models.
	In order to construct the isochrones for the age range
0.1 -- 20\,Gyr that extend to the tip of the RGB, we have constructed 
evolutionary tracks for mass 0.4 -- 5.2 $M_\odot$, from the pre-MS birthline 
to the onset of helium burning.
	In addition to these full isochrones, we also provide
younger isochrones that reach only up to the upper MS, as explained in \S 5.

\subsection{Initial Models at Pre-MS Birthline}

	An important improvement of the $Y^{2}$ isochrones over RYI is
the fact that these stellar models begin, not at the classic ZAMS,
but at the deuterium MS, also called the ``stellar birthline'',
where stars initially become visible objects (Palla and Stahler 1993), during
the pre-MS stage.
	Thus, these models include pre-MS evolution.

	The pre-MS lifetime, denoting the time taken by a star to evolve
from the birthline to the ZAMS, is a strong function of stellar mass, varying
from less than 1\,Myr to about 200\,Myr for stars in our mass range.
	It is about 43\,Myr long for a solar mass model.
	This means that the ZAMS, as hitherto defined, has an inherent age
spread of about 200\,Myr, making it simply inappropriate to use models
starting from the ZAMS to match the HRDs of young clusters.
	Because of this, isochrones that label ZAMS stars as
``zero-age'' are bound to mismatch the lower part of the observed HRDs of
young clusters that contain pre-MS stars (e.g. Patten \& Pavlovsky 1999).
	This effect is negligible in old ($>1$\,Gyr) populations, causing
age underestimates of only 1\% for 10\,Gyr populations, but it can be
significant in the analysis of young open clusters.
	We therefore suggest that
ages be henceforth defined relative to the birthline rather than the ZAMS.

	In view of the expanding utility of isochrones beyond the treatment of
old systems, in recognition of the increasing importance of very young
systems, and to provide a unified treatment for open, globular and
perhaps other star systems, we have decided to include pre-MS
evolution in this set of isochrones.

	This raises the issue of where and how the models should be started
off.
	Is there some equivalent of the ZAMS on the Hayashi track?
	One cannot begin the evolution at an arbitrary point on the pre-MS,
since the radius and the central density of stars, both
important quantities, especially from the viewpoint of stellar rotation,
but also in a more general sense, change rapidly during this phase.
	This deprives us of the pause on the MS, when hydrogen
burning begins, that serves as a natural starting point for models relevant to
old stellar populations.
	An equivalent of this pause is provided by
the deuterium MS, or stellar birthline, as defined by Palla
and Stahler (1991)\footnote{We note that, in contrast to the situation with
the MS, this is a theoretical construct, and there is not, at least
not yet, an observational definition of the deuterium MS.}.
	Its use for these purposes seems both apposite and physically
meaningful.
	For the low-mass stars in this study, the birthline is distinct from
and above the MS, located just above the T Tauri stars in the
HRD.

	We have constructed a series of birthline models for low-mass
stars on the birthline. 
	These models are then evolved forward in time
along the pre-ms tracks, onto the MS and beyond, ending at the
RGB tip, but thus making available isochrones intended for quite young
systems as well, perhaps valid even for 10\,Myr-old systems.

	Operationally, these models were started off from polytropic models
that were placed even higher than the birthline in the HRD.
	However, polytropic models are quite different from real stellar
models in their thermal structure. 
	Thus they need to be relaxed until their thermal structures are
correct.	
	This is accomplished in YREC and the resulting structures are evolved 
down their Hayashi tracks until they satisfy the mass-radius relationship
of Palla \& Stahler (1991), at which point they are assigned age zero (see
also Barnes \& Sofia [1996]). 
	This pre-birthline evolutionary phase is a convenient way of 
calculating physically self consistent initial models, and has no special 
physical significance since it does not include deuterium burning in the
definition of the birthline.  
	Deuterium burning is however implicitly included in the post-birthline
evolution in the reaction network of YREC.  
	The initial deuterium gets destroyed, but the contribution of 
deuterium burning to the total energy production is negligible compared 
to the contribution from gravitational contraction.   
	Thus, the starting models have both the correct, or at least
appropriate, radius and also internal structure\footnote{These issues are
particularly important in studies of the early rotational evolution of stars
because of the importance of the stellar moment of inertia.}.
	The chemical composition is now checked, and adjusted to be pristine.
	Models with differing chemical compositions may be constructed
by rescaling the corresponding solar mixture models. A series of these
models have been constructed for different masses.
	These models are then evolved onto the MS and beyond.

\section{Color Transformations}

	Theoretical properties ([Fe/H], $L$, $T_{\rm eff}$) have been
transformed into colors and magnitudes using the color transformation tables
of Lejeune, Cuisinier, \& Buser (1998, hereafter LCB)
and of Green et al. (1987, hereafter GDK), which was used in the RYI.
	Both tables are semi-empirical: the LCB table is based on the
latest Kurucz spectral library (1992) and the GDK table is based on
the older Kurucz library (1979).
	Both tables have been substantially modified to match the
empirical stellar data better than their forebears (Kurucz tables)
and extended to low
temperatures based on empirical data and low temperature stellar models.
	For the present calculations, we have used the updated {BaSeL}-2.2
version
of the LCB library\footnote{The {BaSeL}-2.2 tables are available only
electronically at {\tt ftp://tangerine.astro.mat.uc.pt/pub/BaSeL/}.},
for which the calibration of the cool giant models in the parameter ranges
4500 K $\leq T_{\rm eff} \leq$ 6000 K and 2.5 $\leq {\rm log}\,g \leq$ 3.5 has
been revised.

	In the process of comparing isochrones to observation, the adoption 
of an up-to-date color transformation table is as
important as updating input physics in the construction of stellar models.
	The difference between the LCB table and the GDK table is substantial,
as shown in Figures 1 and 2.
	Figure 1 shows the comparison of bolometric correction ($BC$)
for ${\rm log}\,g = 4.5$ MS stars.
	The overall agreement is good.
	At solar metallicity and temperature (vertical line in the top right
panel), $BC$ from the LCB table is somewhat higher than that from
the GDK table.
	For example, $BC_{\odot}$ is $-0.08$ and $-0.109$ in the GDK
table and in the LCB table, respectively.
	We have normalized these two tables to the Sun so that the
visual absolute magnitude of the Sun ($M_{V \odot}$) becomes 4.82
(Livingston 2000), regardless of the table used.
	Then, the following definition of $BC$:
\begin{equation}
	BC = M_{bol \odot} - M_V
\end{equation}
yields $M_{bol \odot} = 4.711$ and 4.74 for the LCB table and the GDK table,
respectively.
	Because $M_V$ is defined as:
\begin{equation}
	M_V = M_{bol \odot} - 2.5 {\rm log} L/L_\odot - BC,
\end{equation}
the two tables yield the same $V$ magnitudes.

\placefigure{Fig 1}

	Figure 2 shows how seriously the two color tables differ from
each other.
	The original GDK table reaches only down to the effective
temperature of 2800K and up to 20000K.
	In order to cover the tip of the metal-rich RGB ($Z \geq 0.06$),
we have extrapolated this table down to 2500K when necessary.
	A few of the youngest and most metal-poor isochrones reach above
effective temperature 20000K.
	In that case, we have used the colors from the LCB table.
	This crude remedy did not cause any noticeable discontinuity
in the isochrones.

\placefigure{Fig 2}

	The LCB table is obviously more up-to-date than the GDK table.
	Yet, it is not always our preference.
	In some circumstances, the isochrones with the GDK table
match the empirical data better.
	However, without knowing the true parameters (metallicity,
reddening, and distance modulus), it is difficult to choose one over the
other.
	Thus, we have decided to provide our isochrones in two formats
based on both tables.
	Users are recommended to try both.
	The color transformation is for the filter systems of
$(UBV)_{Johnson}(RI)_{Cousins}$ in the GDK table and of
$(UBV)_{Johnson}(RI)_{Cousins}(JHKLM)_{ESO}$ in the LCB table.

\section{Results}

\subsection{Stellar Evolutionary Tracks}
	Figure 3 shows a sample set of stellar evolutionary tracks for the
solar composition and for the mass range of 0.4 -- 5.0 $M_\odot$.
	The line connecting the diamond symbols is the birthline for this
composition.
	The first position of each track in the HRD is somewhat
uncertain because it depends on the choice of the first time step in the
numerical computation.
	This uncertainty can be alleviated if we force the first time step
to be very small.
	However, we have decided not to pay too much attention to such
details in the
first steps of the pre-MS stage because we believe that other uncertainties
in the position of the birthline and in observational data are much larger.
	In this sense, our youngest isochrones below 10\,Myr are less
certain than their older counterparts.

\placefigure{Fig 3}

\subsection{Isochrones}
	Figure 4 shows a set of isochrones for $Z=0.02$ and ages of
1\,Myr through 20\,Gyr.
	The young isochrones (1, 2, 4, 8, 10, 20, 40, 60, 80 Myr) are complete
only to the MS and thus focused on the pre-MS stage.
	As mentioned earlier, one of the most notable features of the
$Y^2$ Isochrones  is the inclusion of pre-MS stellar evolution.

	The information that we provide in the $Y^2$ Isochrones  is listed in
Table 3.
	The last three columns are luminosity functions (LFs); namely,
the number of stars in the box defined by the mass grid given assuming there
were initially 1000 stars in the mass range of 0.5 -- 1.0 in $M_\odot$,
following the concept introduced in the RYI.
	The bottom right panel in Figure 4 shows the evolution of LF
in the case of IMF slope $x=1.35$ (Salpeter index).

\placefigure{Fig 4}

\placetable{tbl-3}

	Figure 5 compares the $Y^2$ Isochrones  and RYI.
	One can see the substantial change in the stellar models in the top
left panel.
	The other three panels show the isochrones in the observer's HRD.
	The effects of the use of the LCB color transformation table are
clear.
	With a close inspection of the lower MS of some isochrones based on
the LCB table, one will find some discontinuities in $U-B$ and $B-V$ colors
(e.g., lower left panel).
	These are partly real because they are also present in the
empirical data, although with a smaller amplitude.
	The discontinuities in the isochrones get amplified
by the fact that the empirical data are scarce especially in terms
of metallicity and gravity grids.
	Besides, the adjustment (correction) between the Kurucz spectral
library (the foundation of our color tables) and the existing low
temperature libraries made by LCB is not as feasible as we had hoped
for because of the large and apparent difference between them
(At times, the flux difference is as large as an order of magnitude).

\placefigure{Fig 5}

	Similarly, peculiarly curved features on the RGB tip are shown in
the LCB-based isochrones (in $U-B$ and $B-V$ colors).
	These features are caused by the non-monotonic temperature-color
relations that are present in the empirical data.
	Lejeune et al. (1997, Figures 6 \& 14) illustrate these effects.
	Whether the empirical stellar data that show such non-monotonic
temperature-color relations are being correctly interpreted and implemented
in the color transformation table remains to be determined through more
rigorous data sampling.
	Because of the limitation of the adopted calibration procedure by LCB,
the corrected LCB colors still differ from the empirical data by as much
as 0.1\,mag in $U-B$ and $B-V$ (see for instance Figure 14 in LCB).
	In this sense, our LCB-based isochrones are subject to such
uncertainties.
	Readers are referred to Lejeune et al. (1997; 1998) for more
information.
	Despite the magnitude of the disagreement in the colors in this
temperature range (often up to 1\,mag!), this may not be an important
issue in the study of most Milky Way globular clusters, because their
metallicities are too low to be affected by this uncertainty.
	But it is a serious problem in metal-rich stellar populations of
the kind found in the bulges of spiral and elliptical galaxies.

	Part of a sample isochrone is shown in Table 4.
	The tables in their entire form are available electronically from
this paper.
	Minor corrections may occur to the tables after publication.
	The updated versions will be available from the authors upon request
or directly from our WEB 
site\footnote{\tt http://achee.srl.caltech.edu/y2solarmixture.htm}.
	We also provide a FORTRAN interpolation routine
that works for metallicity and age interpolation via the same WEB site.

\placetable{tbl-4}

\subsection{Bump on the Red Giant Branch}

	The evolution in luminosity on the RGB is not always a monotonic
function of time.
	When the depth of the convective envelope increases, some
processed material is distributed over the entire convective region,
and CNO processed isotopes are mixed to the surface.
	Stars are rejuvenated by such mixing because unprocessed
hydrogen is mixed from the envelope into the interior.
	The first, and largest, of these partial mixing phenomena on the RGB,
is called ``first dredge up'', an event which takes place when the star's
convection zone reaches its maximum depth soon after it reaches the giant
branch.
	The convection zone depth then decreases (in terms of mass fraction)
as the star evolves up the giant branch, leaving behind a composition
discontinuity.
	At some later point, the hydrogen burning shell passes through the
composition discontinuity.
	This event gives rise to a temporary decrease in luminosity,
as shown in Figure 6.
	This was predicted by Thomas (1967).
	This results in a peak in the luminosity function on the RGB,
or ``bump'', which is observed in some globular clusters
(see e.g. King et al. 1985; Fusi Pecci et al. 1990).
	At luminosities higher than the bump on the giant branch, stars will
have more fuel to burn in the upper part of the RGB, because the first
dredge-up had previously mixed in fresh hydrogen from the envelope.

\placefigure{Fig 6}

	Figure 7 and Table 5 display the predicted position of the RGB bump
as a function of age and metallicity.
	Except when the metallicity is extremely low, the RGB bump generally
gets fainter as age and metallicity increase.
	We have also plotted in Figure 7-(b) the $V$ magnitudes of the
RGB-bumps in some globular clusters.
	The $V$ brightness data are from Table 4 in Fusi Pecci et al.,
while the distance moduli and the metallicities are from Harris (1996).
	In the case of 47\,Tuc (the second most metal-rich data point), we have
adopted a distance modulus of 13.47 from our CMD fit described in 
Figure 13.
	The most metal-rich data point ($V(bump) \approx 17.1$) 
has been derived from the NGC\,6553 data from Zoccali et al. (2001).
	Adopting [Fe/H]=$-$0.17 (see \S 7.3), we derived 
$M_{V}(bump) \approx 1.3$, which is also in good agreement with the
prediction.

	The prediction of the position of the RGB bump cannot be precise 
because of the uncertainties in the detailed physics at the base of the 
convection zone and the dynamics of the first dredge-up. 
	When the RGB bump was first identified in 47\,Tuc (King et al. 1985) 
the observed RGB bump luminosity then differed from the predicted
luminosity in the theoretical tracks, and possible causes for this 
discrepancy were explored, principally  convective  overshoot below
the convection zone, and opacity uncertainties.  
	Subsequent improvements in the quality of the observational data and
in the opacities and other physics input have resulted in a decrease in
the discrepancy (Fusi Pecci et al. 1990).  
	More recently, some researchers have argued that the discrepancy 
disappears when one takes into account the opacity increase due to
alpha-enhancement and that the  effect of convective overshoot is negligible 
(Cassisi and Salaris 1997; Ferraro et al. 1999). 
  
	We note at this point that other physical processes can play an 
equally significant role. 
	The presence of mixing other than convective overshoot complicates 
the issue.
	There is evidence that some mixing below the convection zone
takes place in red giants, probably the result of meridional circulation
due to rotation (Sweigart \& Mengel 1979).  
	Near the MS, the envelope overshoot does not seem sufficient to 
explain the observed lithium depletion below the convective envelope 
(Pinsonneault 1994).  
	We also note that helioseismological inversions indicate that there 
is little overshoot below the convection zone in the present sun 
(Basu, Antia, \& Narasimha 1994). 
	For all these reasons, our models were calculated without any 
overshoot below the convection envelope.  
	The effects of including alpha-element enhancement on the RGB bump 
position will be discussed in our next paper.

\placefigure{Fig 7}

\placetable{tbl-5}

\subsection{Maximum Brightness of the RGB}

	In general, the RYI did not provide accurate positions of
RGB tips, because until recently isochrones were dominantly used
for HRD fitting, a matter in which the RGB tip plays a small, if any, role.
	However, isochrones are now used in a variety of studies.
	One example is a field of evolutionary population synthesis,
whose aims include the computation of
integrated properties (eg. colors and magnitudes) of stellar populations.
	The position of the RGB tip is important in computing near infrared
magnitudes, as stars near the RGB tip are good infrared sources.
	Therefore, we have paid extra attention in order to define the RGB tip
as consistently as possible.
	It should be noted, however, that evolution near the RGB tip
remains somewhat uncertain, and will remain so as long as mass loss is
poorly understood.

	Another important application of accurately modeled RGBs is
to use them as distance indicators for metal-rich populations.
	Figure 8 shows that the brightness in $V$ magnitude
on the RGB reaches a maximum not at the RGB tip but in the middle of the RGB.
	Because the number density of stars in the middle of the RGB is
much higher than that of the RGB tip, this gives us a large statistical
advantage.
	Besides, this part is more certain than near the RGB tip
in terms of our understanding of stellar evolution.
	This effect, however, occurs only when the metallicity is high enough,
i.e., $Z \gtrsim 0.004$, because it is mainly an opacity effect.
	There are currently a great number of
observational studies that are based on this concept.
	New predictions are provided in Figure 8 and Table 6.

\placefigure{Fig 8}

\placetable{tbl-6}

\section{Luminosity Functions and Photometric Evolution}

	Isochrones are convenient building blocks for
the the study of evolutionary population synthesis (EPS).
	EPS studies either use the luminosity functions (LFs) provided
in the isochrone tables (if provided as is the case with our isochrones)
or compute the LFs using the mass information that is always provided in the
isochrones.
	Generating isochrones is not a trivial matter partly because
keeping the desired accuracy in mass through several track interpolations
is not easy.
	As Yi et al. (2000) mentioned, the mass difference between
the bottom and the tip of the RGB in the 12\,Gyr isochrones is merely an order
of 0.7\% or approximately 0.006 $M_\odot$.
	If the mass interpolation is inaccurate in the 5th decimal place
in $M_\odot$, the LF will lose its desired accuracy.
	Such errors can be checked by inspecting the integrated photometric
properties of a sample population, as shown in Figure 9.
	Our isochrones all show clear and smooth luminosity (and color)
evolution, except where there are expected departures from
the monotonicity in the color transformation table.
	Considering this complexity, it is better to use the isochrones
rather than stellar tracks in most of the EPS studies.

\placefigure{Fig 9}

\section{Tests of Isochrones against Empirical Data}

	We are encouraged by the fact that the improved isochrones actually
provide a better match to the empirical data.
	In this section, we display some of the tests we have made.
	We have attempted to provide a variety of tests to check the validity
of different aspects of the isochrones.
	For example, we try to test our isochrones on young open
clusters as well as old globular clusters.
	We demonstrate the level of agreement between the isochrones and
the MS stars in the field as well as in an open cluster.
	We also present a test of the color evolution of Galactic
globular clusters.
	We are pleased to find that our isochrones combined with their
LFs reproduce their integrated colors quite well at their generally
accepted ages.

\subsection{Population I MS stars}

	One of the most basic tests is to examine whether our stellar models
match the measured properties of nearby MS stars.
	Figure 10 shows a set of stellar evolutionary tracks for $Z=0.02$,
i.e., nearly solar.
	Overplotted are the observed MS stars from Gray (1992).
	The match is good: not only do the models match the locus of the
observed MS, but the match in mass is also good except in the very low
mass range ($M \lesssim 0.6$ $M_\odot$).
	The masses of the faintest three data points (M dwarfs) are
0.6, 0.52, and 0.48 $M_\odot$ according to Gray (1992), while models suggest
approximately 0.7, 0.6, 0.58 $M_\odot$.
	This disagreement in mass in the low mass range may be due to
the uncertainties in the mass determination of cool stars or in the
low-temperature stellar atmosphere models.

\placefigure{Fig 10}

\subsection{Subdwarf MS Stars}

	Figure 11 shows the match between our isochrones and the subdwarf MS
stars whose distance has been determined through HIPPARCOS observations.
	The data are from Reid (1998), and we have followed the same
criteria for the data selection as VandenBerg et al. (2000) did for their
Figure 14: i.e., $[Fe/H]\,<\,-0.55$, $M_{V}\,>\,4.5$,
and $\sigma_{\pi}/\pi\,<\,0.07$.
	Instead of 14\,Gyr isochrones, we are using 12\,Gyr isochrones.
	However, there is practically no difference in this part of the
isochrones.
	Since our isochrones do not include the effects of $\alpha$-element
enhancement, we have recalibrated to compute the isochrones in this plot
using Equation [2].
	We have adopted [$\alpha$/Fe]=+0.4 and +0.2 for [Fe/H]$\leq-$1.0 and
[Fe/H]=$-$0.5, respectively, following Wheeler, Sneden, Truran (1989) and
Carney (1996).
	The agreement is good and is marginally better with the isochrones
based on the GDK table.

\placefigure{Fig 11}

\subsection{Globular clusters}

	Globular clusters are the main targets of isochrone applications.
	They are the playgrounds of stars of the same age and metallicity,
so we can test and refine our stellar models.
	Moreover, an accurate estimation of their ages has been appreciated
as an important route to a better understanding of cosmology.

	Figures 12 and 13 shows fits to the metal-rich cluster 47\,Tuc and
the metal-poor cluster M\,68, using our isochrones based on the GDK table.
	We have used theoretical values of $M_{V}$(RR) from
Demarque et al. (2000) to derive their distances.
        These models are based on the HB evolutionary tracks with input physics
nearly identical to the current isochrones.
        They also take into account the effects of HB morphology in the
calculation of $M_{V}$(RR) for a given [Fe/H].
        The theoretical values of $M_{V}$(RR) from these models are in a
reasonable agreement with RR Lyrae luminosities derived from the main
sequence fitting of globular clusters based on HIPPARCOS parallaxes for
field subdwarfs (\cite{r97}; \cite{gra97}).
        For purposes of illustration, we have shown synthetic HB models in
the lower panels (see Lee, Demarque, \& Zinn 1994 for description).

        In case of 47\,Tuc, its red HB is approximately 0.15 mag brighter
than RR Lyrae level.
        The matches are good if their ages are approximately 12 and 13\,Gyr,
respectively.
	Through the tests on a few other globular clusters, we have found a
reduction in the age estimates of the order of 20\%, and this is when
$\alpha$-enhancement has been considered using the Salaris et al. formula.

\placefigure{Fig 12}

\placefigure{Fig 13}

	The primary source of this age reduction is the update of the
stellar models.
	Of several ways to measure the ages of globular clusters, the 
most popular one is probably to match the MSTO luminosity using isochrones. 
	The $Y^2$ isochrones are fainter at the MSTO (bluest point on the MS) 
than the RYI by approximately 16\% when $Z=0.0004$, a metallicity
typical for Galactic globular clusters. 
	Figure 14 clearly shows this. 
	A given observed MSTO luminosity indicates substantially younger 
ages when the $Y^2$ isochrones are used.
	Also compared in this figure is the result derived from the isochrones
of Girardi et al. (2000), whose input physics is quite compatible to ours.
	The agreement with our models is good.

\placetable{tbl-7}

\placefigure{Fig 14}

	Table 7 shows the change of the derived ages caused by the
use of the new stellar models alone.
	The estimates are the mean  values of the age estimates for various 
values of MSTO luminosities within the range of 5 -- 16 Gyrs. 
	Much of this age reduction is caused by the inclusion of the the
helium-diffusion and the use of the updated equations of states.
	Our preliminary investigation suggests that, in addition to this, 
the inclusion of alpha-enhancement causes approximately another 5\% of 
age reduction. 
	It should also be noted that the change (reduction) in derived ages 
caused by the improvement in the stellar models alone (no alpha-enhancement 
effects included) is reversed in the metal-rich regime (see Table 7).
	These issues will be discussed more thoroughly in our next paper.

	The next test is on NGC\,6553, a metal-rich Galactic globular
cluster (Figure 15).
	The faint data (dots) are from Zoccali et al. (2001).
	They are confirmed members of the cluster.
	Also shown are the bright stars ($V<17$: crosses) from
Sagar et al. (1999).
	They have been selected to satisfy $Error(V-I)<0.2$.
	These two data sets have offsets both in $V$ magnitude and 
in $V$-$I$ by 0.3 mag and 0.15 mag (the Zoccali et al. data being brighter 
and bluer), respectively.
	This is because the Zoccali et al. data are from the region that
suffers less reddening (Zoccali, priv. comm.).
	When we shifted the Sagar et al. data, both sets match one another
from the MS to the HB clump very well.
	The magnitudes of the offsets were estimated from the mean positions
of the HB clump through eye-fit.

	The metallicity of this cluster has been reported to be 
in the range of [Fe/H]=$-$0.5 -- +0.1 (see Sagar et al. 1999), i.e. 
approximately $Z \approx$ 0.006 -- 0.02.
	We have adopted $Z$=0.0125 ([Fe/H]=$-$0.17) in this study because 
at this metallicity the LCB-based $Y^2$ isochrones match the upper RGB 
of this cluster best.
	The age of the model (9\,Gyr for this fit, $E$($V$-$I$)=0.9, 
$m$-$M$[apparent]=15.78) has been derived from the luminosity difference 
between the MSTO and the mean luminosity of the HB clump.
	The open diamond shown on top of the HB clump is the synthetic
HB clump for this age and metallicity.
	However, the age estimate is sensitive to the adopted metallicity;
10 -- 8\,Gyr when $Z$=0.01 -- 0.02.
	Zoccali et al. (2001) achieve a somewhat larger age estimate, 12\,Gyr,
but this is mainly because they assume a metallicity lower than ours.
	Using LCB-based $Y^2$ isochrones, we favor our metallicity and
thus a smaller age.

	The match by the LCB-based $Y^2$ Isochrones is good.
	The turn-around in the $V$ magnitude would not be as dramatic as 
observed if the GDK table is used, although GDK-based isochrones
seem to match the MS better.
	Considering the importance of this bulge cluster to the study
of the formation of Milky Way, 	a more careful analysis, in particular
considering $\alpha$-enhancement (highly suspected to be present 
[B. Barbuy, priv. comm.], is required to give a definite answer.

\placefigure{Fig 15}

\subsection{Open clusters}

	If tests against globular clusters teach us about the stellar
evolution in the MS and phases beyond, those against young open clusters teach
us about pre-MS stellar evolution.
	Several studies have noted that the theoretical isochrones
are fainter in the lower MS than observed (e.g., Subramaniam \& Sagar 1999;
Patten \& Pavlovsky 1999).
	This is likely to be evidence of pre-MS evolution.

	Figures 16 and 17 show isochrone fits to the young open clusters,
Pleiades and IC\,2391.
	Both LCB- and GDK-based isochrones work well for these fittings.
	The Pleiades data are very well represented by the 40--100\,Myr
isochrones, assuming $E(B-V)=0.04$ and $m-M$(apparent)=5.6
(Pinsonneault et al. 1998).
	A good fit requires an apparent distance modulus somewhat
larger than that derived from the HIPPARCOS data, as was noted earlier
by Pinsonneault et al. (1998).

\placefigure{Fig 16}

\placefigure{Fig 17}

	The overall fit appears to be good, but the data at
$B-V \approx 0.5 $ and $\gtrsim 1.5$ seem to be notably brighter (or redder)
than the models.
	If the observational errors are not to be considered large,
this may indicate the shortcomings of the models.

	The MS data of the younger cluster IC\,2391 is matched by the 20\,Myr
solar-composition isochrone.
	This is reasonably close, if somewhat smaller than the age estimate
of Patten \& Pavlovsky (1999), i.e. 30\,Myr.
	The membership study of this cluster has been performed very
carefully, and thus the data are of good quality and the sequence is better
defined than the data shown in Figure 16.
	The model isochrones are for 4, 10, 20, 40, and 80\,Myr of age,
while the 20\,Myr isochrone is marked as thick line.
	The 20\,Myr isochrone appears to be matching the data reasonably well,
but this pre-MS isochrone does not fully reproduce the curvatures at
$V-I \approx 2$.
	A similar but larger mismatch was reported by Patten \& Pavlovsky
(1999; see their Figure 3).
	It seems that the $Y^2$ Isochrones  match the data better in the
lower MS (cross symbols in Figure 16) than the fit shown in their Figure 3,
which was based on a different set of isochrones.

	The level of agreement between the model and the data may not yet be
satisfying to some readers.
	However, we are quite impressed by the matches, considering the
large uncertainties still embedded in the convection approximation in
cool stars and their color transformation.
	Fits to open clusters involve complex considerations of membership, 
photometry of differing quality for stars in differing magnitude ranges, 
(differential) reddening, metallicity, and of course, distance, all of which
impinge on the age assigned. 
	Inconsistencies are often found and resolved only in the context of 
detailed studies, and we invite the community to use these isochrones in such 
contexts.

\subsection{Photometric Evolution}

	One of the most exciting outcomes of our isochrone tests is
that the integrated colors can be practically used as reliable age indicators
of simple stellar populations.
	Astronomers have long been aware of this fact, but the level of
the accuracy of integrated colors as age indicators has never been very high.
	Optical integrated colors were believed to be useful in
determining whether a population is 5\,Gyr or 15\,Gyr, but not whether
it is 10\,Gyr or 13\,Gyr; that it, the expected accuracy was of an order of
50\%.
	This is because the color evolution of old populations is
too slow to be useful as an age indicator.
	In this paper, we claim that we have achieved a substantially
better precision.
	Figure 18 shows the mean integrated colors of Galactic
globular clusters (Harris 1996) compared with the models.
	Thin lines are the models made of only MS and RGB stars and
thick lines are the complete models with all phases including HB stars.
	When complete models are used, the mean colors of the Milky Way
clusters indicate ages of 12 -- 13\,Gyr, which is virtually identical
to the age estimates obtained from the isochrone fitting.
	If this is not a coincidence, this improvement in the predictability
of integrated colors as a function of age and metallicity will be useful
to many applications in the population study.

\placefigure{Fig 18}

\section{Conclusions}

	If the HRD opened the window into the details of stellar evolution,
it is isochrones that made the knowledge of stellar evolution practically 
meaningful to other areas of astronomy.

	In this first paper, we publish a set of isochrones in which the 
heavy elements are scaled to the solar mixture.   
	The use of the solar mixture is quite appropriate for the study of
many stellar populations with approximately solar relative abundances. 
	Isochrones to the tip of the giant branch for ages ranging from 
0.1 to 20 Gyr have been constructed for metallicities $Z$ between 0.00001 
and 0.08. 
	Additional isochrones with ages down to 1 Myr are also provided for 
the pre-MS and the MS phases only.

	A follow-up paper will consider mixtures that are enhanced in  
$\alpha$-elements as compared to the sun, since many pieces of evidence 
suggest $\alpha$-element enhancement in metal-poor stars and some metal-rich 
populations.
	Meanwhile, the reader is provided with a convenient correction 
formula for $\alpha$-element enhancement.  
	This formula has proved reliable in the case of metal-poor stars in
the field and in globular clusters which are believed to be $\alpha$-element 
enhanced.

	Important advantages of the $Y^2$ Isochrones  over the RYI are as
follows.
	(1) The $Y^2$ Isochrones  are based on updated input physics.
	(2) They are available for the more up-to-date color transformation
table of Lejeune et al. (1998), as well as for the Green et al. table
used for the RYI.
	(3) The stellar models start their evolution from the pre-MS
birthline instead of from ZAMS.
	This makes the $Y^2$ Isochrones  the first comprehensive isochrones
(covering a large range of age and metallicity) based on the stellar models
that include the pre-MS stage.
	This fact makes the $Y^2$ Isochrones  useful to young open
cluster studies as well.
	(4) Convective core overshoot has been taken into account.
This has important impacts on isochrone fitting to open cluster HRDs
and to the spectral dating of intermediate-age populations.

	We note that the $Y^2$ Isochrones match the properties of observed 
objects much better than the RYI.   
	In particular, good matches are achieved at substantially 
smaller ages than before for the Galactic globular clusters, 
by approximately 20\%, once $\alpha$-element enhancement corrections are
taken into account. 
	Approximately 15\% (out of 20\%) of this age reduction is caused by 
the update in the stellar models.

	Partly owing to the improved color calibration, and partly due to 
the inclusion of post-RGB evolutionary phases, the model integrated colors 
generated by the isochrones now also match within the uncertainties 
the observed integrated colors of Galactic globular clusters at their
expected ages.
	This is one of the most significant achievements made by the $Y^2$
Isochrones.

	The $Y^2$ isochrones are available from the authors upon request
or directly from our WEB 
site\footnote{\tt http://achee.srl.caltech.edu/y2solarmixture.htm}, 
as well as electronically through this paper.
	We are currently constructing a new set of $\alpha$-element enhanced
$Y^2$ Isochrones.

\acknowledgements

	We thank the referee for useful comments and constructive questions.
	We thank Ata Sarajedini, Brian Patten, Annapurni Subramaniam,
Eva Grebel, Manuela Zoccali, and Alistair Walker for making the cluster 
HRDs available to us.
	We are particularly grateful to Guillermo Torres for finding
an error in an early version of the $Y^2$ isochrones.
	We also thank Beatriz Barbuy, Sergio Ortolani and Leo Girardi for 
useful comments on isochrones.
	This work was in part supported by the Creative Research Initiative
Program of the Korean Ministry of Science \& Technology grant (SY, YK, YL).
	PD acknowledges support from NASA Grants NAG5-8406 and NAG5-6404.
	TL gratefully  acknowledges  financial support    from the
``Funda{\c{c}}\~ao para a  Ci\^encia e Tecnologia'' (Portugal), (grant
PRAXIS-XXI$/$BPD$/22061/99$, and grant  PESO/P/PRO/15 128/1999).
	SB would like to acknowledge the McKinney Foundation and the NSF for
support under grants AST-9731302 and AST-9986962 to the Univ. of Wisconsin.

\begin{table*}
\caption{Input Physics and Parameters} \label{tbl-1}
\begin{center}
\begin{tabular}{cc}
\tableline
\tableline
Solar mixture			& Grevesse \& Noels (1993)\\
OPAL Rosseland mean opacities	& Rogers \& Iglesias (1995), Iglesais \& Rogers (1996) \\
Low temperature opacities	& Alexander \& Ferguson (1994) \\
Equations of state		& OPAL EOS (Rogers et al. 1996) \\
Energy generation rates		& Bahcall \& Pinsonneault (1992; 1994 priv. comm.) \\
Neutrino losses			& Itoh et al. (1989) \\
Convective core overshoot 	& 0.2 $H_{p}$ for age $\leq$ 2\,Gyr\\
Alpha-element enhancement 	& none \\
Helium diffusion 		& Thoul et al. (1994) \\
Mixing length parameter		& ${\it l}/H_{p} = 1.7431$ \\
%Eddington T-$\tau$ relation	& \\
Primordial helium abundance 	& $Y = 0.23$ \\
Galactic helium enrichment parameter & $\Delta$$Y$/$\Delta$$Z = 2.0$ \\
\tableline
\tableline
\end{tabular}
\end{center}
\end{table*}

\begin{table*}
\caption{Chemical Compositions and Convective Core Developing Mass}\label{tbl-2}
\begin{center}
\begin{tabular}{rrrr}
\tableline
\tableline
$Z$	& $Y$		&	[Fe/H]	&	$M^{conv}_{crit}$ \\
\tableline
0.00001	& 0.23002	&	-3.29	&	2.1 \\
0.00010	& 0.23020	&	-2.29	&	1.6 \\
0.00040	& 0.23080	&	-1.69	&	1.4 \\
0.00100	& 0.23200	&	-1.29	&	1.3 \\
0.00400	& 0.23800	&	-0.68	&	1.2 \\
0.00700	& 0.24400	&	-0.43	&	1.2 \\
0.01000	& 0.25000	&	-0.27	&	1.2 \\
0.02000	& 0.27000	&	 0.05	&	1.2 \\
0.04000	& 0.31000	&	 0.39	&	1.1 \\
0.06000	& 0.35000	&	 0.60	&	1.1 \\
0.08000	& 0.39000	&	 0.78	&	1.0 \\
\tableline
\tableline
\end{tabular}
\end{center}
\end{table*}

\begin{table*}
\caption{Description of the  $Y^2$ Isochrones} \label{tbl-3}
\begin{center}
\begin{tabular}{cc}
\tableline
\tableline
Column 	&	Description \\
\tableline
1	& Mass in $M_\odot$ \\
2	& ${\rm log} T$ \\
3	& ${\rm log} L/L_{\odot}$ \\
4	& ${\rm log} g$ \\
5	& $M_{V}$ \\
6 	& $U-B$ \\
7 	& $B-V$ \\
8 	& $V-R$ \\
9	& $V-I$ \\
10 	& $V-J$\tablenotemark{a} \\
11 	& $V-H$\tablenotemark{a} \\
12 	& $V-K$\tablenotemark{a} \\
13 	& Star number density for power law IMF index $x=-1$\tablenotemark{b} \\
14 	& Star number density for power law IMF index $x=1.35$\tablenotemark{b}\\
15 	& Star number density for power law IMF index $x=3$\tablenotemark{b} \\
\tableline
\tableline
\end{tabular}
\end{center}
\tablenotetext{a}{This color is not available in the isochrones based on the GDK table.}
\tablenotetext{b}{Number of stars in the box defined by this mass grid where
there are 1000 stars in the IMF in the mass range of 0.5 -- 1.0 $M_\odot$.}
\end{table*}

\begin{table*}
\caption{Part of a sample isochrone. The whole isochornes are available electronically.}\label{tbl-4}
\begin{center}
\begin{tabular}{rrrrrrrrrrrrrrrrr}
\tableline
\tableline
 M/Msun  &  logT & logL/Ls  & logg &  Mv  &  U-B &  B-V &  V-R &  V-I  &  V-J  &  V-H &  V-K &  V-L &  V-M  &  N(x=-1)  &  N(x=1.35)  &  N(x=3) \\
\tableline
 0.4000000& 3.5645& -1.6026 &4.8526&10.051 &1.076& 1.506& 1.030& 1.982& 3.701& 3.486& 3.701& 3.701& 4.701& 1.9551E+01& 7.1323E+01& 1.5597E+02 \\
 0.4195508& 3.5800& -1.5126& 4.8453& 9.613& 1.031& 1.418& 0.934& 1.790& 3.429& 3.223& 3.429& 3.429& 3.429& 3.9115E+01& 1.3132E+02& 2.7100E+02 \\
 0.4391149& 3.5956& -1.4223& 4.8372& 9.198& 0.960& 1.323& 0.860& 1.656& 3.136& 2.938& 3.136& 3.136& 3.136& 3.9099E+01& 1.1793E+02& 2.2573E+02 \\
 0.4586498& 3.6113& -1.3322& 4.8288& 8.830& 0.876& 1.239& 0.775& 1.501& 2.938& 2.754& 2.938& 2.938& 2.938& 3.9141E+01& 1.0656E+02& 1.8980E+02 \\
 0.4782554& 3.6269& -1.2416& 4.8188& 8.492& 0.784& 1.165& 0.686& 1.336& 2.797& 2.635& 2.797& 2.797& 2.797& 3.9170E+01& 9.6656E+01& 1.6068E+02 \\
 0.4978195& 3.6425& -1.1510& 4.8080& 8.181& 0.719& 1.102& 0.626& 1.216& 2.691& 2.543& 2.691& 2.691& 2.691& 4.0790E+01& 9.1419E+01& 1.4204E+02 \\
 0.5190451& 3.6578& -1.0594& 4.7957& 7.867& 0.644& 1.036& 0.570& 1.103& 2.570& 2.434& 2.570& 2.570& 2.570& 4.3124E+01& 8.7724E+01& 1.2734E+02 \\
 0.5409435& 3.6730& -0.9681& 4.7832& 7.535& 0.532& 0.955& 0.507& 0.982& 2.389& 2.264& 2.389& 2.389& 2.389& 4.4283E+01& 8.1763E+01& 1.1088E+02 \\
 0.5633278& 3.6882& -0.8773& 4.7708& 7.233& 0.379& 0.875& 0.466& 0.912& 2.210& 2.095& 2.210& 2.210& 2.210& 4.5107E+01& 7.5726E+01& 9.6053E+01 \\
 0.5860503& 3.7033& -0.7869& 4.7580& 6.960& 0.213& 0.797& 0.438& 0.868& 2.063& 1.957& 2.063& 2.063& 2.063& 2.6800E+01& 4.1776E+01& 5.0287E+01 \\
 0.5901278& 3.7061& -0.7709& 4.7562& 6.916& 0.184& 0.782& 0.435& 0.862& 2.043& 1.939& 2.043& 2.043& 2.043& 8.1951E+00& 1.2332E+01& 1.4479E+01 \\
 0.5942454& 3.7088& -0.7547& 4.7538& 6.871& 0.157& 0.768& 0.431& 0.856& 2.025& 1.922& 2.025& 2.025& 2.025& 8.3248E+00& 1.2324E+01& 1.4304E+01 \\
 0.5984526& 3.7115& -0.7383& 4.7513& 6.827& 0.131& 0.754& 0.428& 0.850& 2.007& 1.906& 2.007& 2.007& 2.007& 8.8200E+00& 1.2838E+01& 1.4725E+01 \\
\tableline
\tableline
\end{tabular}
\end{center}
\end{table*}

\begin{table*}
\caption{Position of the Red Giant Branch Bump}\label{tbl-5}
\begin{center}
\begin{tabular}{rrrrrrrrr}
\tableline
\tableline
[Fe/H] & Age(Gyr) & ${\rm log} T$ & ${\rm log} L/L_{\odot}$ & Mv & U-B & B-V & V-R & V-I \\
\tableline
-2.29 &  2 &  3.692 &  2.618 & -1.467 &  0.083 &  0.746 &  0.487 &  0.972\\
-2.29 &  5 &  3.692 &  2.441 & -1.024 &  0.076 &  0.744 &  0.486 &  0.972\\
-2.29 &  7 &  3.692 &  2.382 & -0.878 &  0.075 &  0.744 &  0.486 &  0.972\\
-2.29 & 10 &  3.692 &  2.321 & -0.724 &  0.076 &  0.746 &  0.487 &  0.973\\
-2.29 & 13 &  3.692 &  2.269 & -0.593 &  0.074 &  0.746 &  0.487 &  0.973\\
-1.29 &  2 &  3.690 &  2.330 & -0.769 &  0.248 &  0.828 &  0.489 &  0.962\\
-1.29 &  5 &  3.689 &  2.141 & -0.295 &  0.245 &  0.827 &  0.489 &  0.962\\
-1.29 &  7 &  3.689 &  2.075 & -0.130 &  0.248 &  0.829 &  0.490 &  0.963\\
-1.29 & 10 &  3.689 &  2.002 &  0.053 &  0.248 &  0.830 &  0.490 &  0.963\\
-1.29 & 13 &  3.688 &  1.946 &  0.194 &  0.249 &  0.831 &  0.490 &  0.964\\
-0.68 &  2 &  3.677 &  2.071 & -0.093 &  0.541 &  0.940 &  0.515 &  1.001\\
-0.68 &  5 &  3.675 &  1.892 &  0.360 &  0.551 &  0.945 &  0.518 &  1.007\\
-0.68 &  7 &  3.675 &  1.826 &  0.529 &  0.556 &  0.947 &  0.520 &  1.010\\
-0.68 & 10 &  3.674 &  1.749 &  0.726 &  0.563 &  0.950 &  0.522 &  1.013\\
-0.68 & 13 &  3.673 &  1.690 &  0.875 &  0.567 &  0.952 &  0.522 &  1.015\\
-0.27 &  2 &  3.665 &  1.913 &  0.355 &  0.822 &  1.039 &  0.557 &  1.046\\
-0.27 &  5 &  3.664 &  1.724 &  0.838 &  0.831 &  1.042 &  0.561 &  1.053\\
-0.27 &  7 &  3.662 &  1.649 &  1.032 &  0.839 &  1.046 &  0.565 &  1.059\\
-0.27 & 10 &  3.661 &  1.577 &  1.219 &  0.847 &  1.049 &  0.568 &  1.066\\
-0.27 & 13 &  3.661 &  1.522 &  1.361 &  0.850 &  1.050 &  0.569 &  1.068\\
 0.05 &  2 &  3.649 &  1.929 &  0.429 &  1.154 &  1.167 &  0.628 &  1.155\\
 0.05 &  5 &  3.653 &  1.622 &  1.169 &  1.089 &  1.134 &  0.613 &  1.129\\
 0.05 &  7 &  3.653 &  1.551 &  1.352 &  1.093 &  1.137 &  0.616 &  1.135\\
 0.05 & 10 &  3.651 &  1.481 &  1.539 &  1.101 &  1.141 &  0.622 &  1.144\\
 0.05 & 13 &  3.650 &  1.421 &  1.697 &  1.108 &  1.144 &  0.627 &  1.152\\
 0.39 &  2 &  3.637 &  1.890 &  0.658 &  1.506 &  1.295 &  0.722 &  1.297\\
 0.39 &  5 &  3.643 &  1.532 &  1.500 &  1.405 &  1.247 &  0.692 &  1.244\\
 0.39 &  7 &  3.642 &  1.458 &  1.695 &  1.409 &  1.249 &  0.698 &  1.253\\
 0.39 & 10 &  3.641 &  1.374 &  1.915 &  1.412 &  1.252 &  0.705 &  1.265\\
 0.39 & 13 &  3.640 &  1.307 &  2.087 &  1.410 &  1.251 &  0.708 &  1.270\\
 0.60 &  2 &  3.628 &  1.902 &  0.741 &  1.729 &  1.383 &  0.774 &  1.402\\
 0.60 &  5 &  3.636 &  1.523 &  1.599 &  1.612 &  1.324 &  0.741 &  1.325\\
 0.60 &  7 &  3.636 &  1.448 &  1.793 &  1.608 &  1.323 &  0.745 &  1.331\\
 0.60 & 10 &  3.636 &  1.368 &  1.998 &  1.603 &  1.323 &  0.747 &  1.336\\
 0.60 & 13 &  3.636 &  1.293 &  2.184 &  1.592 &  1.318 &  0.747 &  1.336\\
 0.78 &  2 &  3.622 &  1.908 &  0.807 &  1.849 &  1.439 &  0.793 &  1.460\\
 0.78 &  5 &  3.630 &  1.550 &  1.611 &  1.757 &  1.387 &  0.774 &  1.397\\
 0.78 &  7 &  3.631 &  1.456 &  1.841 &  1.742 &  1.381 &  0.773 &  1.395\\
 0.78 & 10 &  3.631 &  1.365 &  2.069 &  1.729 &  1.377 &  0.773 &  1.394\\
 0.78 & 13 &  3.631 &  1.297 &  2.237 &  1.718 &  1.373 &  0.772 &  1.393\\
\tableline
\tableline
\end{tabular}
\end{center}
\end{table*}

\begin{table*}
\caption{Maximum brightness ($M_{V}$) of the RGB as a Function of Age and Metallicity ($Z$)}\label{tbl-6}
\begin{center}
\begin{tabular}{rrrrrrrrrrrrrr}
\tableline
\tableline
t(Gyr)& 0.00001 & 0.0001 & 0.0004 & 0.001 & 0.004 & 0.007 & 0.01 & 0.02 & 0.04 & 0.06 & 0.08 \\
\tableline
 2   & -1.737& -2.392& -2.671& -2.584& -2.207& -1.919& -1.573& -1.002& -0.415& -0.052& 0.231\\
 3   & -2.091& -2.597& -2.703& -2.635& -2.099& -1.727& -1.369& -0.789& -0.185& 0.179& 0.457\\
 4   & -2.423& -2.673& -2.708& -2.568& -1.982& -1.586& -1.228& -0.645& -0.044& 0.340& 0.622\\
 5   & -2.528& -2.709& -2.706& -2.543& -1.889& -1.486& -1.123& -0.538& 0.083& 0.463& 0.749\\
 6   & -2.573& -2.730& -2.702& -2.522& -1.787& -1.399& -1.036& -0.453& 0.193& 0.564& 0.852\\
 7   & -2.600& -2.744& -2.699& -2.503& -1.720& -1.319& -0.953& -0.372& 0.271& 0.657& 0.939\\
 8   & -2.618& -2.756& -2.698& -2.488& -1.655& -1.253& -0.883& -0.304& 0.335& 0.733& 1.015\\
 9   & -2.637& -2.768& -2.698& -2.473& -1.594& -1.202& -0.833& -0.242& 0.393& 0.799& 1.082\\
10   & -2.644& -2.770& -2.694& -2.456& -1.550& -1.150& -0.784& -0.195& 0.459& 0.839& 1.139\\
11   & -2.649& -2.772& -2.689& -2.441& -1.509& -1.111& -0.745& -0.148& 0.510& 0.878& 1.183\\
12   & -2.654& -2.773& -2.684& -2.426& -1.469& -1.071& -0.703& -0.113& 0.563& 0.912& 1.217\\
13   & -2.659& -2.774& -2.678& -2.410& -1.441& -1.039& -0.669& -0.070& 0.597& 0.960& 1.247\\
14   & -2.663& -2.775& -2.674& -2.394& -1.403& -1.004& -0.634& -0.039& 0.639& 1.007& 1.277\\
15   & -2.666& -2.775& -2.670& -2.383& -1.376& -0.974& -0.607& -0.003& 0.672& 1.055& 1.325\\
\tableline
\tableline
\end{tabular}
\end{center}
\end{table*}

\begin{table*}
\caption{Change of age estimate caused by the update in the stellar models alone} \label{tbl-7}
\begin{center}
\begin{tabular}{cc}
\tableline
\tableline
 Z & Age Change (\%) \\
\tableline
0.0001  & -17.21 \\
0.0004 	& -15.87 \\
0.001   & -17.16 \\
0.004   & -10.57 \\
0.02    &  10.02 \\
\tableline
\tableline
\end{tabular}
\end{center}
\end{table*}

{}

%\end{document}

\begin{figure}
\plotone{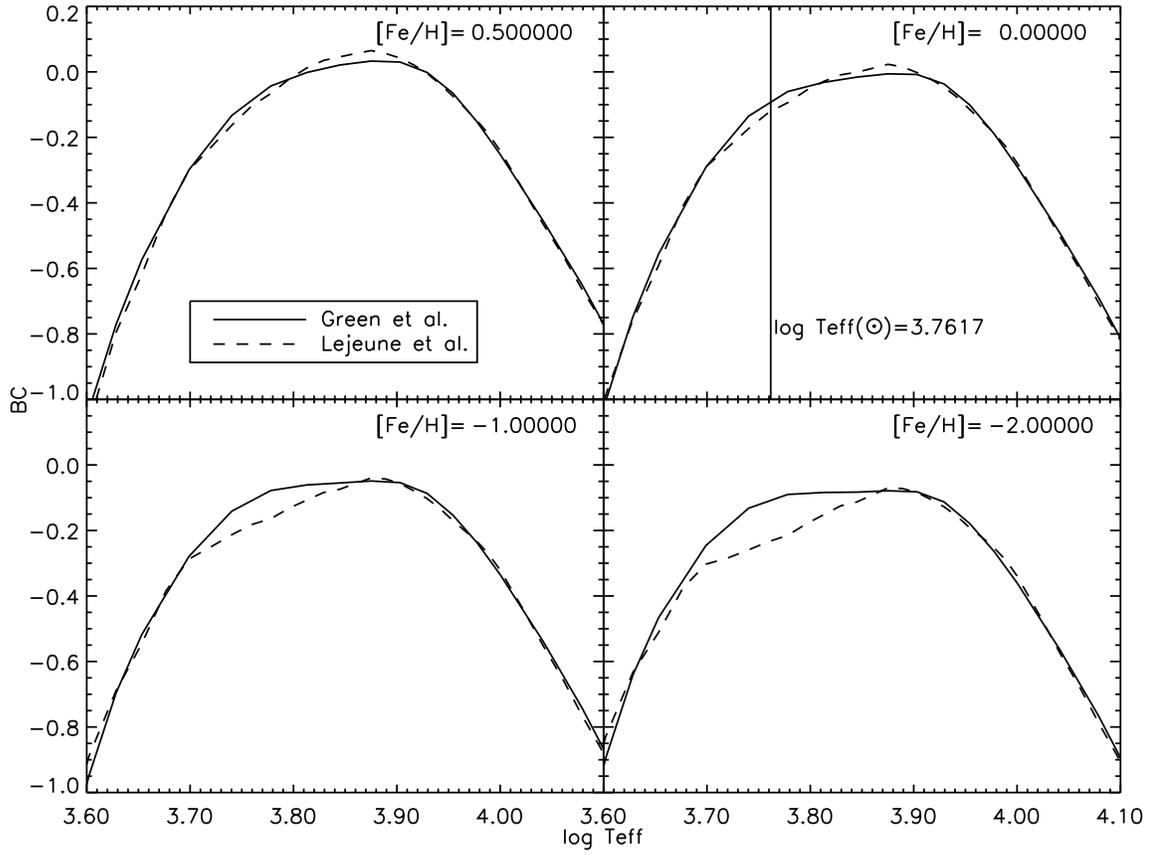}
\caption{Comparison of bolometric correction (BC). The comparison is
for the stars of ${\rm log}\,g = 4.5$. The difference is larger in
metal-poor stars.
\label{fig1}}
\end{figure}

\begin{figure}
\plotone{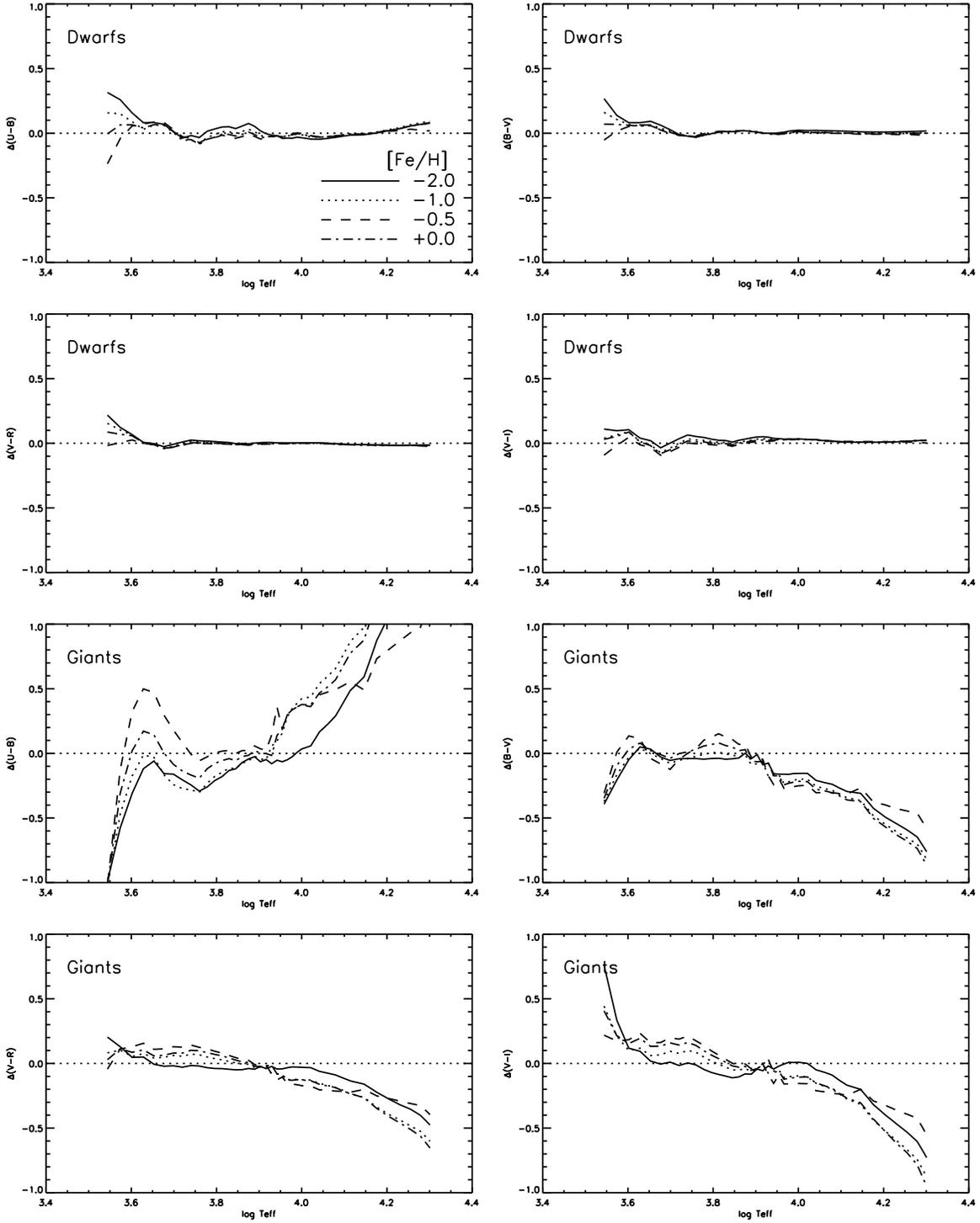}
\caption{Comparison in colors between the LCB table and the GDK table.
The y-axis shows LCB color minus GDK color.
Dwarfs are for ${\rm log}\,g = 4.5$ and Giants are for ${\rm log}\,g = 0.5$.
The difference is substantial for giants.
\label{fig2}}
\end{figure}

\begin{figure}
\plotone{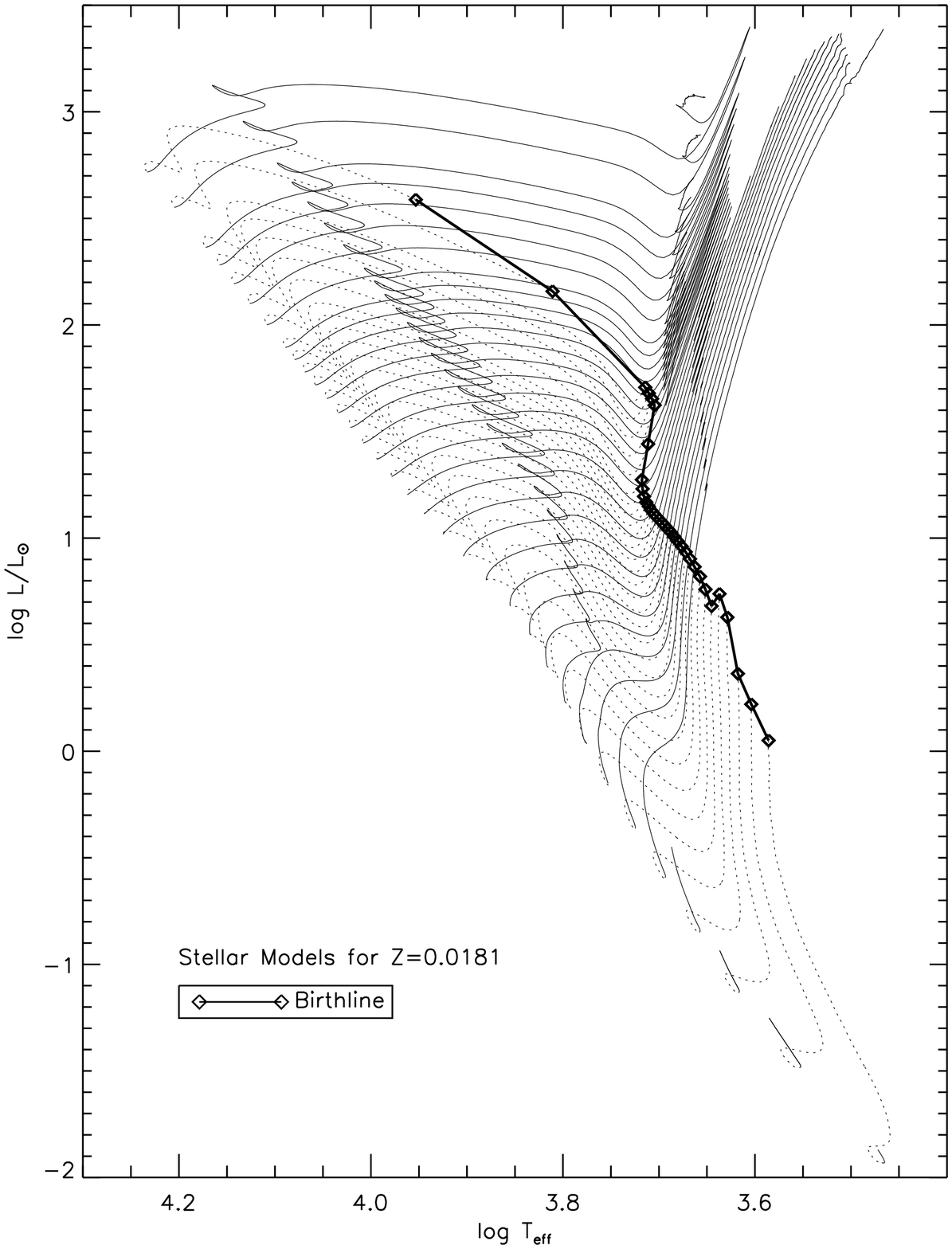}
\caption{Stellar evolutionary tracks from the pre-MS birthline to the RGB,
all for the solar composition. The pre-MS phase is shown in dotted lines.
\label{fig3}}
\end{figure}

\begin{figure}
\plotone{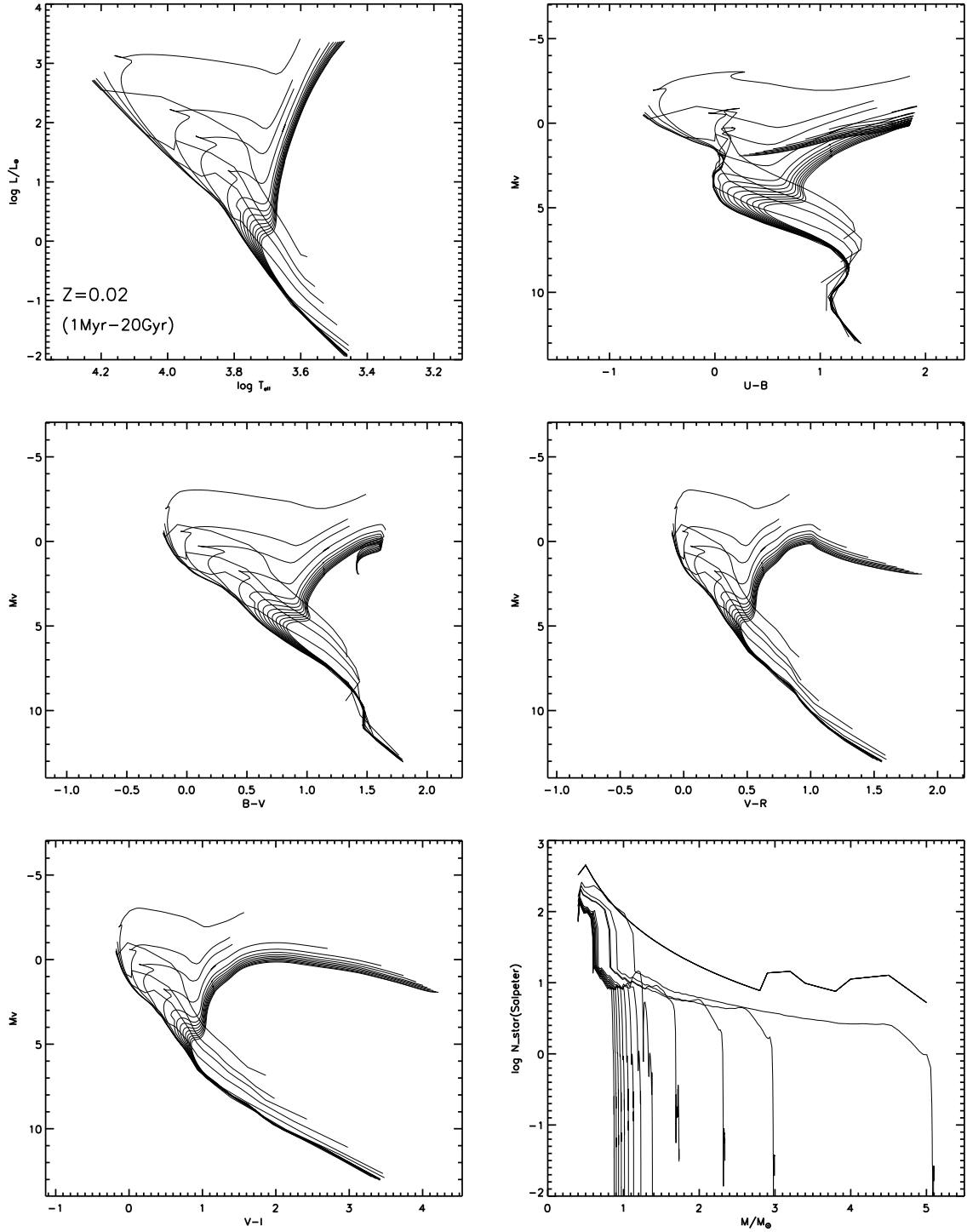}
\caption{A set of sample isochrones for $Z=0.02$, based on the LCB table .
The bottom right panel shows the luminosity functions for all isochrones
but the youngest ones of age smaller than 0.1\,Gyr, where
the farthest to the right (top) is the youngest (0.1\,Gyr).
\label{fig4}}
\end{figure}

\begin{figure}
\plotone{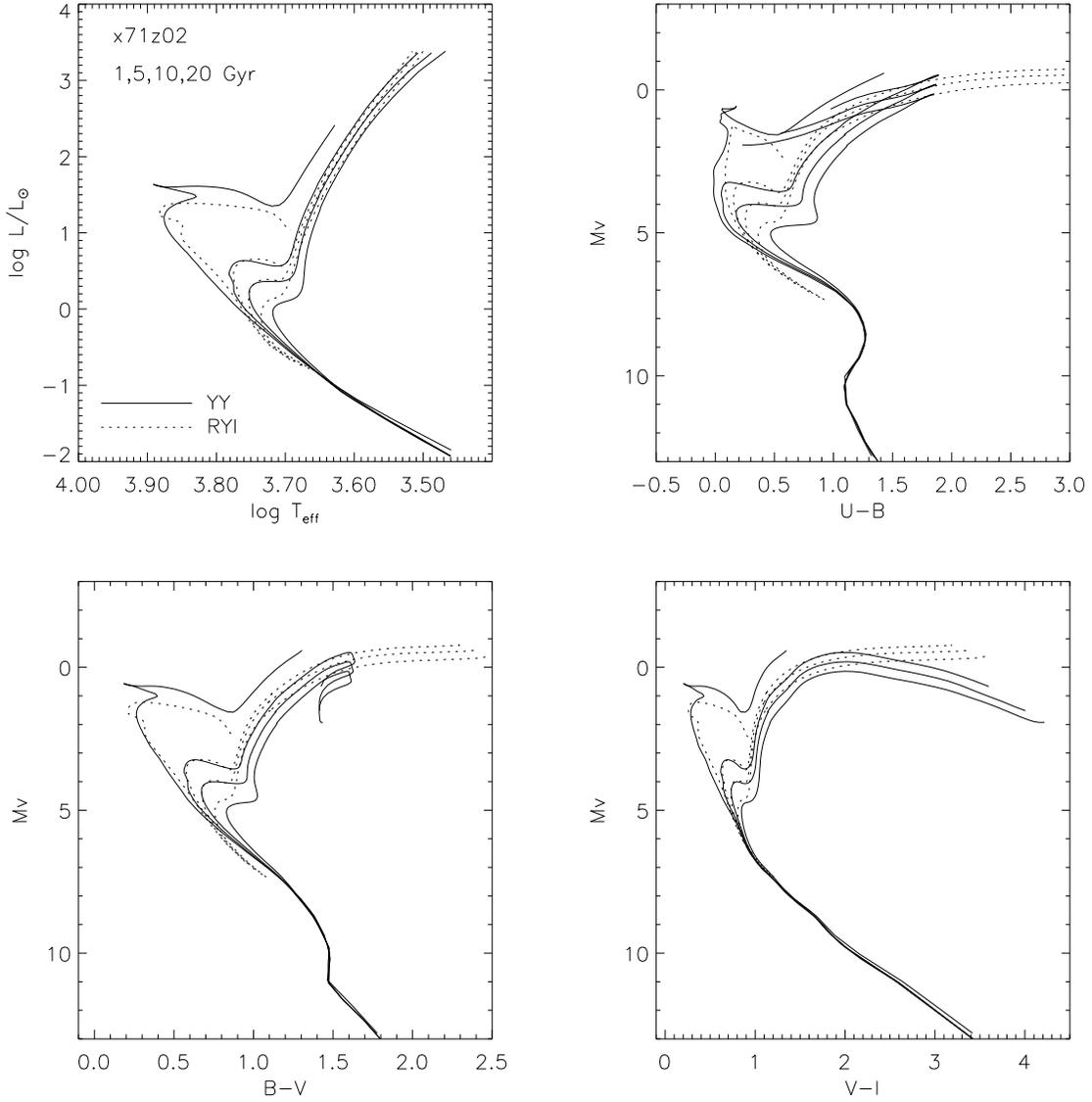}
\caption{Comparison between the Revised Yale Isochrones and the new isochrones
(based on the LCB table) for $Z=0.02$. The top left panel shows the
difference in theoretical HRD which is caused by the difference in stellar
models. The other panels show the effects of the use of the updated color
transformation table.
\label{fig5}}
\end{figure}

\begin{figure}
\plotone{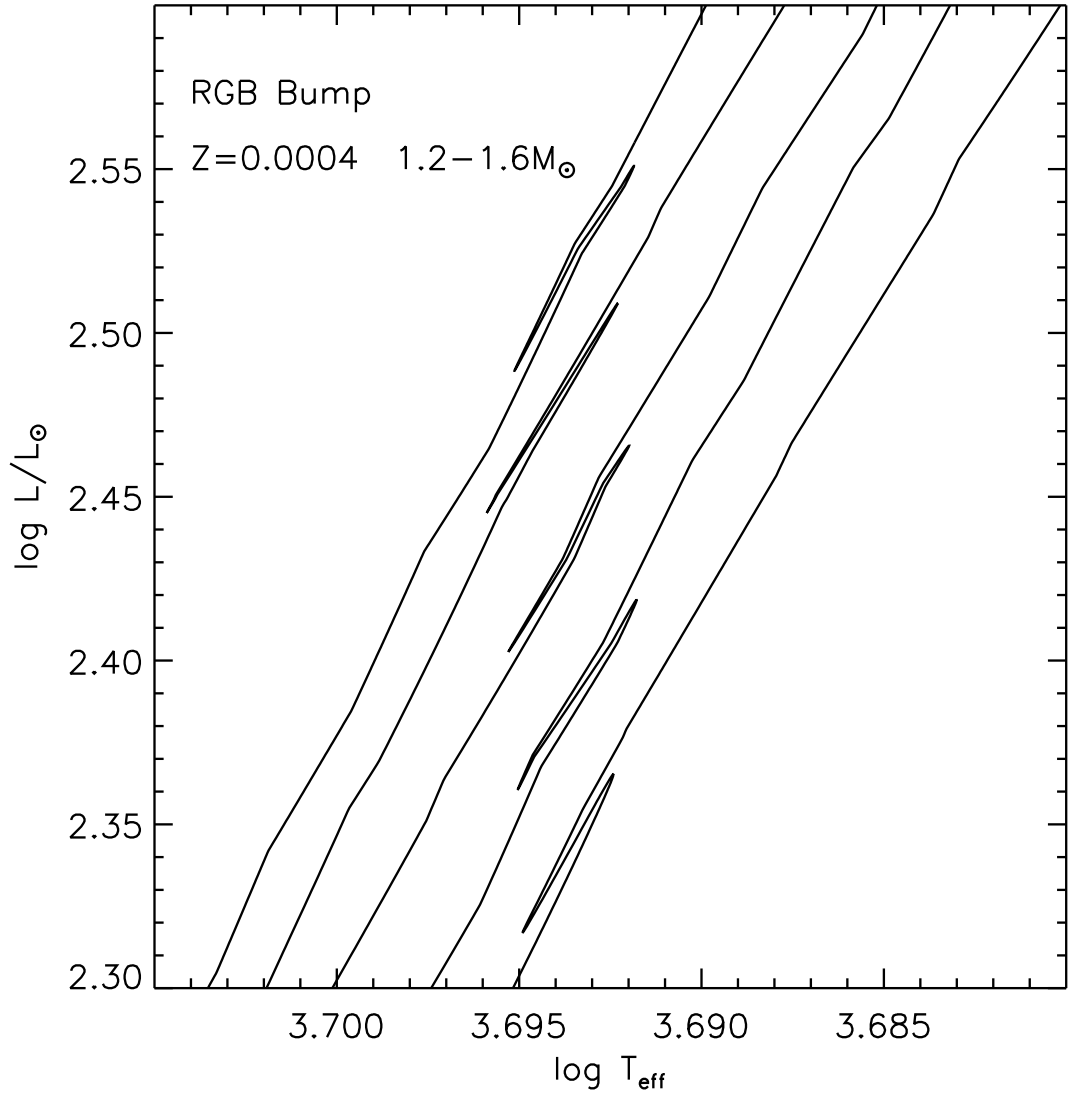}
\caption{Metal poor RGB tracks that show first dredge up bump phenomenon
(see text).
>From bottom to top, models are for 1.2, 1.3, 1.4, 1.5, 1.6 $M_\odot$.
The models shown here are for no convective core overshoot.
\label{fig6}}
\end{figure}

\begin{figure}
\plotone{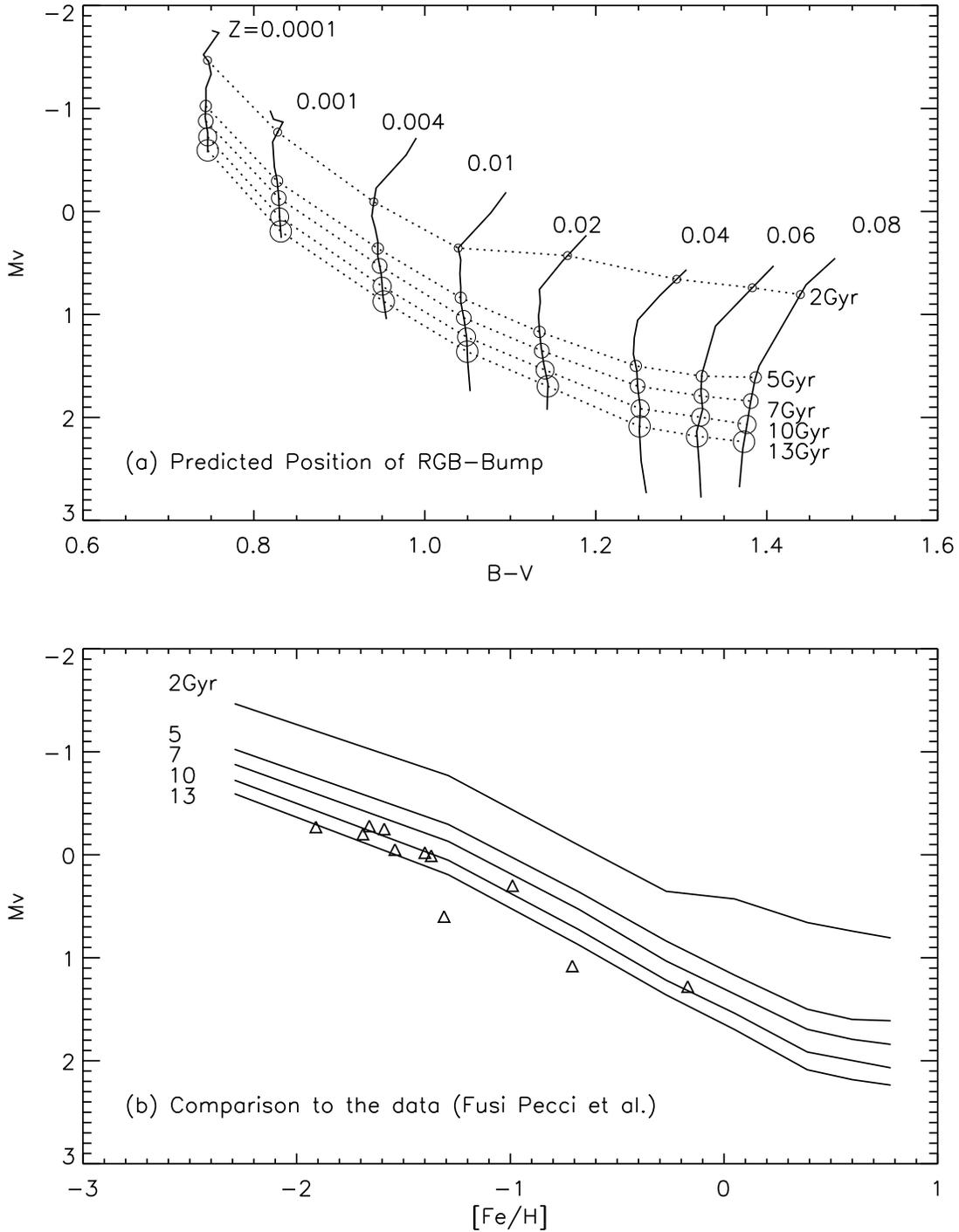}
\caption{The position of the RGB bump caused by the first dredge-up,
as a function of age and metallicity (based on the LCB table).
(a) Each sequence is denoted by its metallicity. Circles are for 2, 5, 7, 10,
and 13\,Gyr of age from the smallest symbol to the largest.
(b) Each line is a metallicity sequence for given age, as denoted in the
far left. Triangles are the $V$ magnitudes of the RGB-Bump adopted
from Fusi Pecci et al. (1990). 
The most metal-rich data point is for NGC\,6553.
The data for this cluster have been kindly provided by M. Zoccali prior
to publication.
\label{fig7}}
\end{figure}

\begin{figure}
\plotone{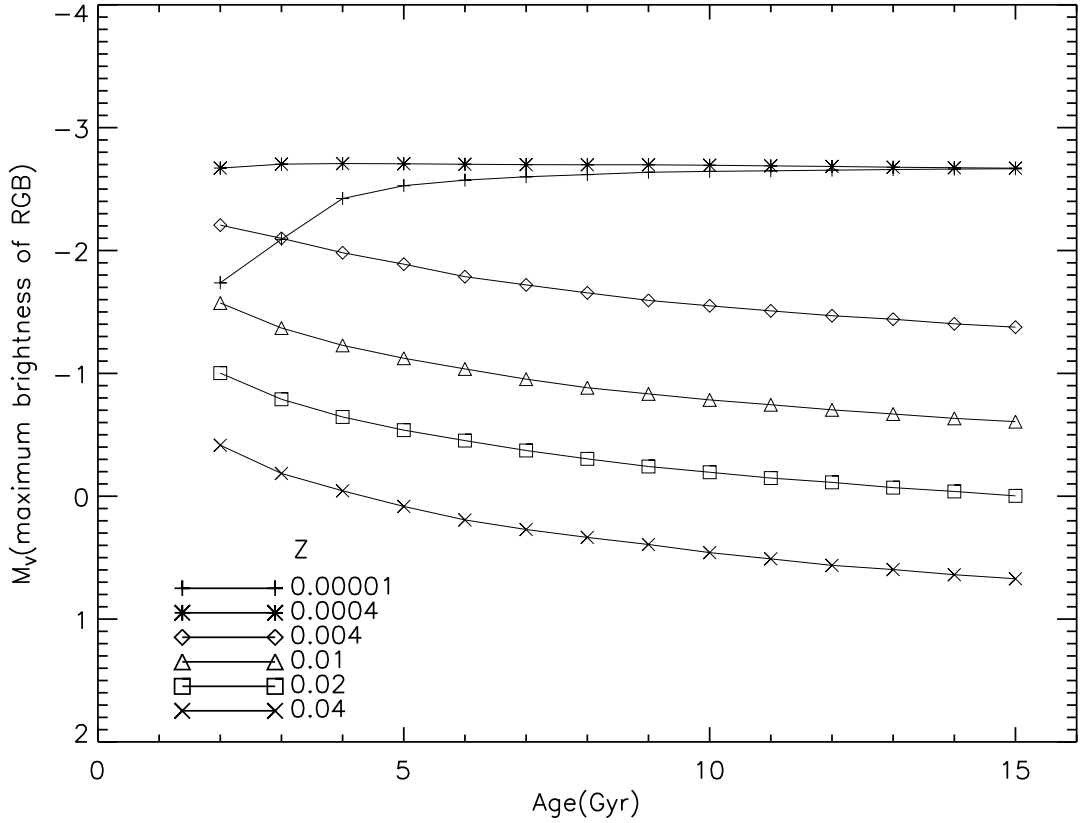}
\caption{Maximum brightest in V magnitude of the RGB as a function of age and
metallicity. The brightest RGB means the tip of the RGB when metallicity is
low ($Z \lesssim 0.004$) but a mid-point on the RGB when metallicity is
high. The LBC table has been used for this because it reaches farther down
to the lower effective temperature than the GDK table does.
\label{fig8}}
\end{figure}

\begin{figure}
\plotone{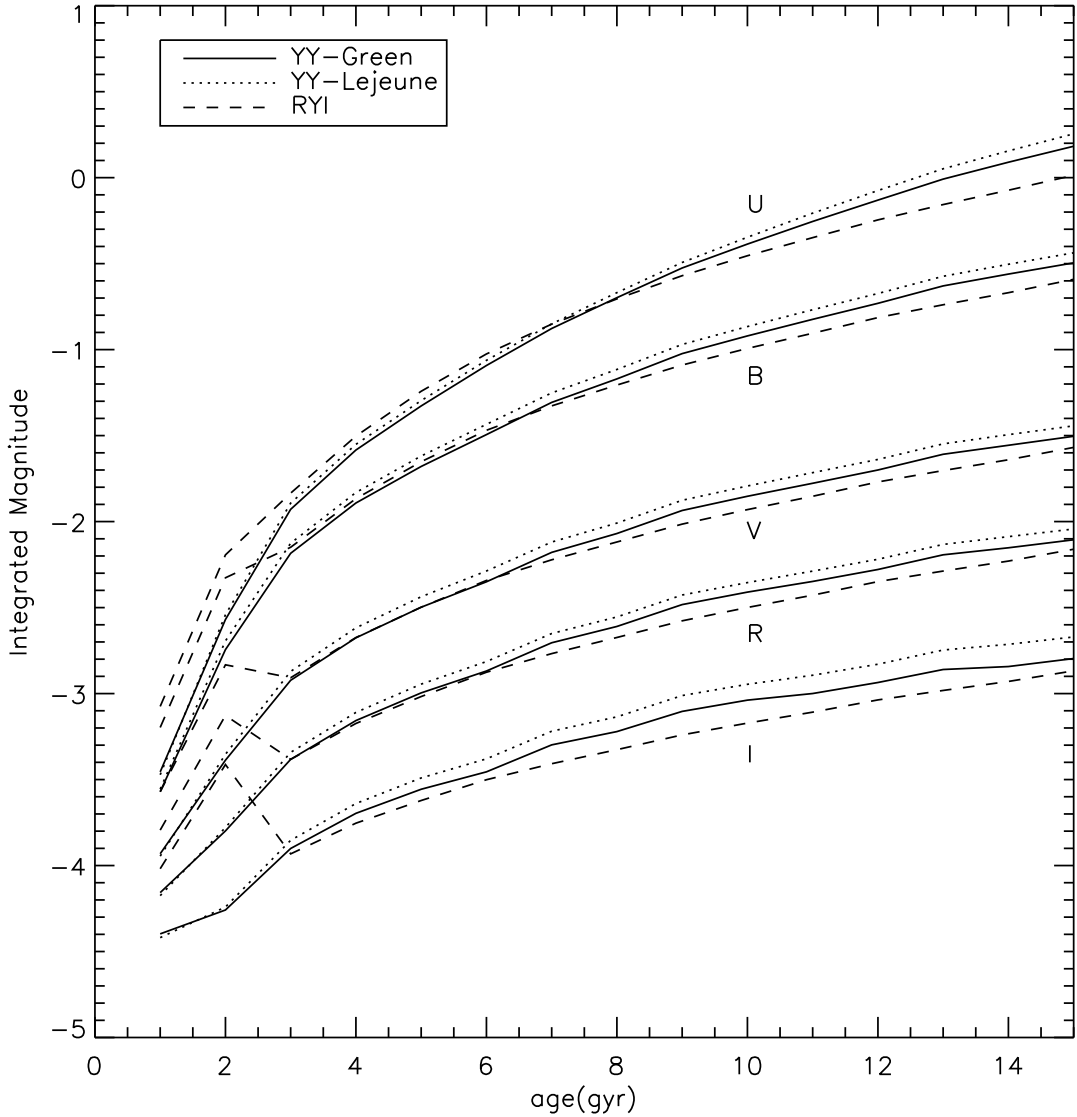}
\caption{The evolution of integrated luminosity for $Z=0.02$ and total mass of
approximately 940 $M_\odot$ in the range of 0.4 -- 1.0 $M_\odot$.
The smoothness of the luminosity evolution is a direct check on the
accuracy of the mass interpolation and thus on the LF.
No matter how accurately the shape of an isochrone may be determined,
LF can easily be in error if the mass interpolation is performed
inaccurately. Such isochrones cannot be used in the evolutionary
population synthesis.
\label{fig9}}
\end{figure}

\begin{figure}
\plotone{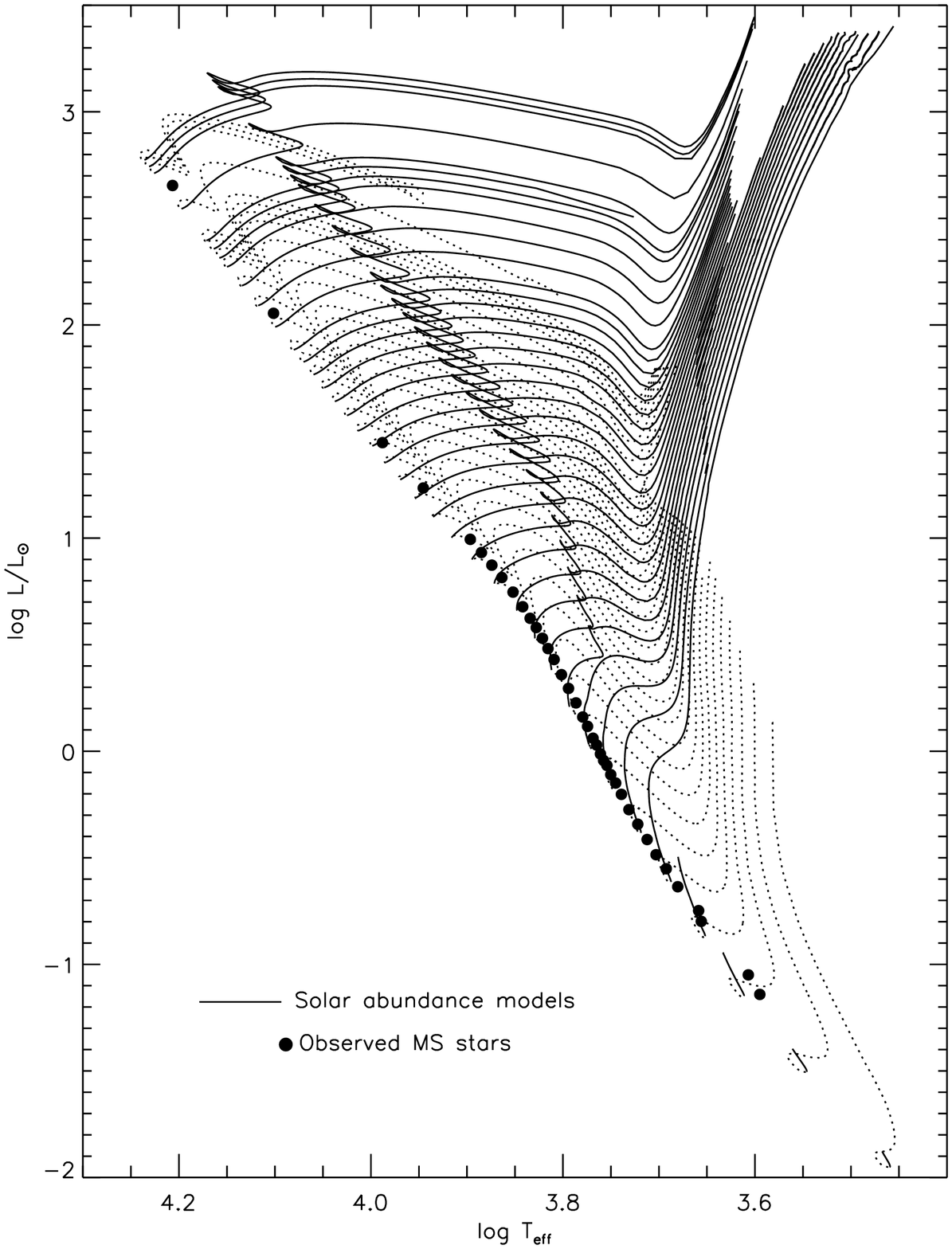}
\caption{Stellar evolutionary tracks from the pre-MS birthline to the RGB,
all for the solar composition. The pre-MS phase is shown in dotted lines.
Filled circles are the observed MS (Gray 1992).
\label{fig10}}
\end{figure}

\begin{figure}
\plotone{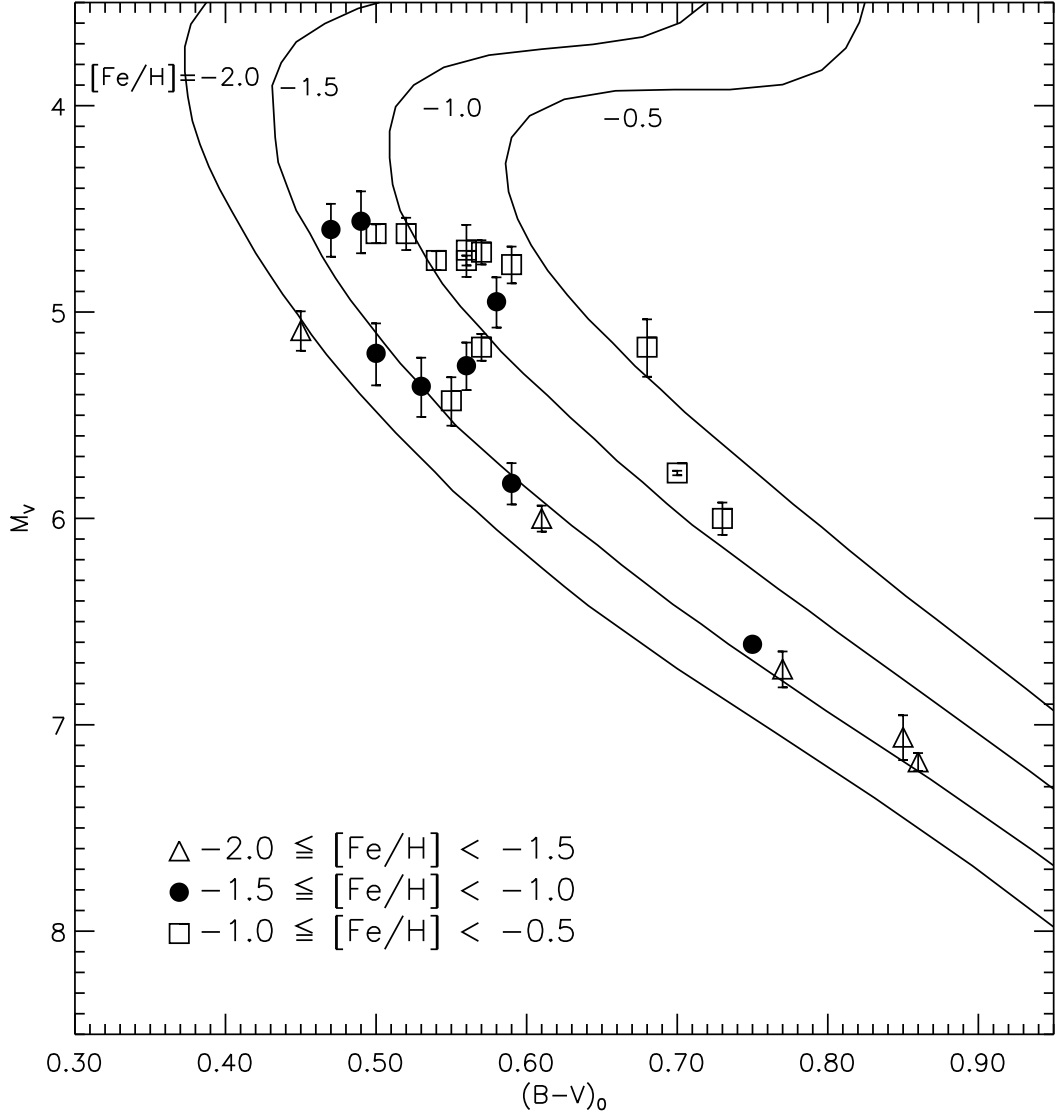}
\caption{A test to the subdwarf MS stars whose distance has been determined
by HIPPARCOS observation. We have used the GDK-based isochrones with
the following $\alpha$-enhancement assumption:
[$\alpha$/Fe]=+0.4 for [Fe/H]$\leq-$1.0, and +0.2 for [Fe/H]=$-$0.5.
\label{fig11}}
\end{figure}

\begin{figure}
\epsscale{0.6}
\plotone{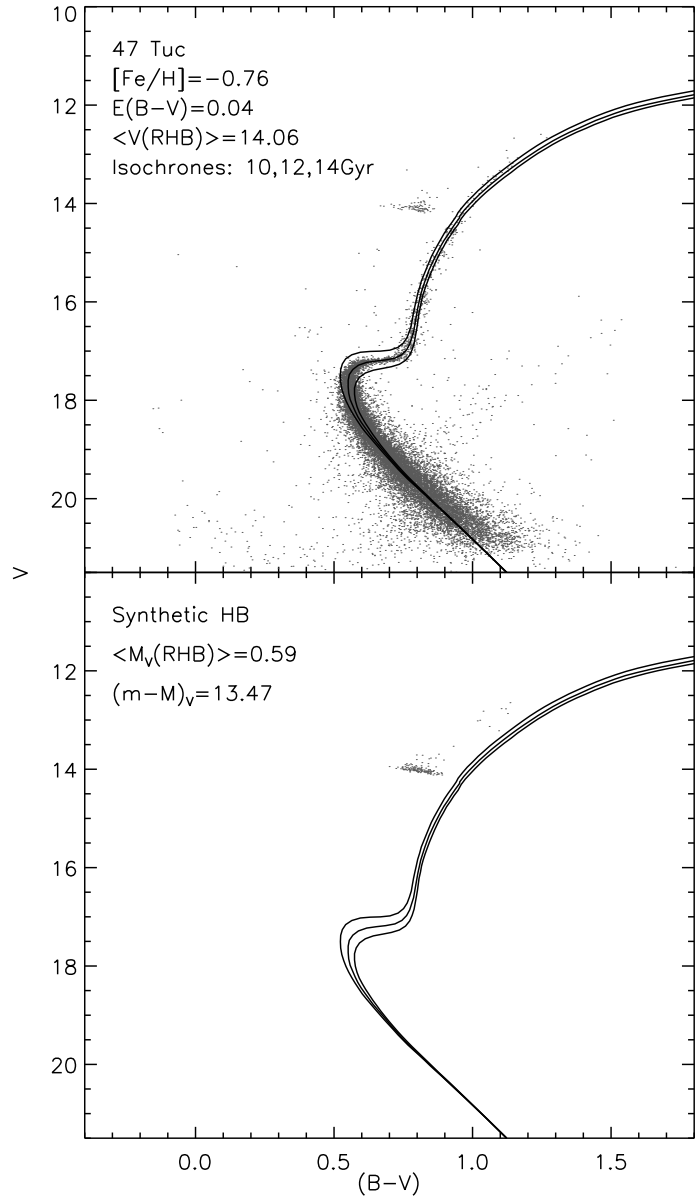}
\caption{A test to the HRD of 47\,Tuc.
The HRD data have been kindly provided by A. Sarajedini.
The estimates of [Fe/H] and reddening are from Harris (1996).
[$\alpha$/Fe]=0.15 adopted following the general pattern in the halo.
\label{fig12}}
\end{figure}

\begin{figure}
\epsscale{0.6}
\plotone{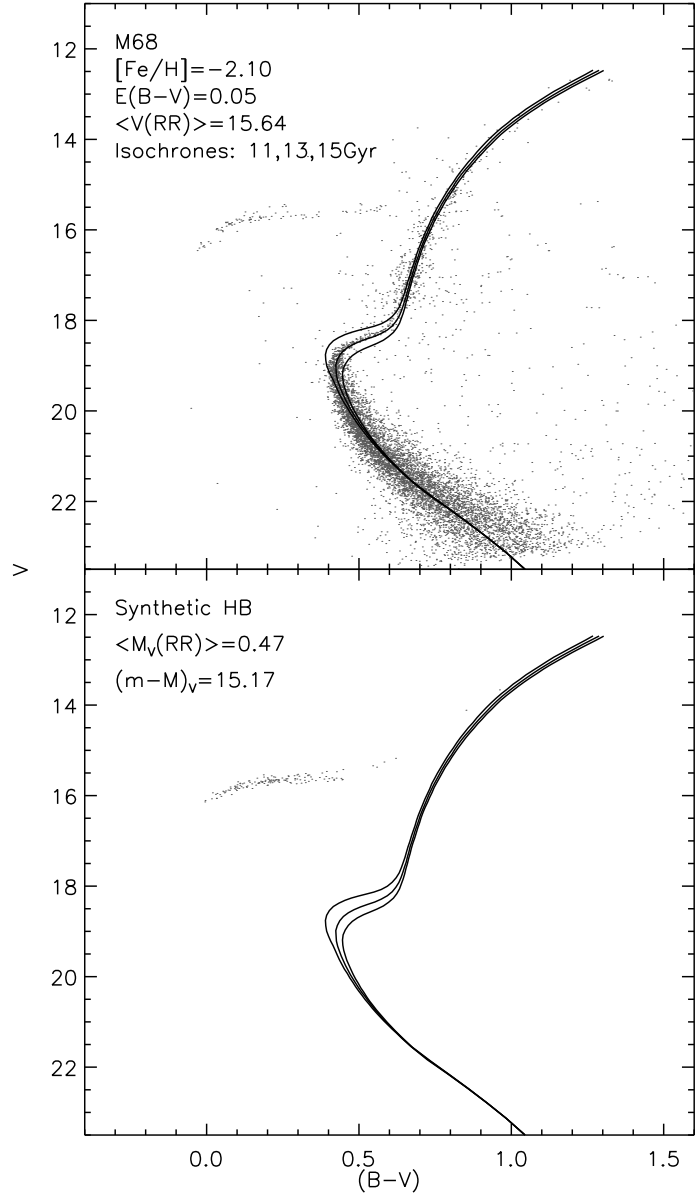}
\caption{Same as Figure 12 but for M\,68.
The HRD data are from Walker (1994).
The estimates of [Fe/H] and reddening are from Harris (1996), and
[$\alpha$/Fe]=0.3 has been adopted following the halo observations.
\label{fig13}}
\end{figure}

\begin{figure}
\epsscale{0.8}
\plotone{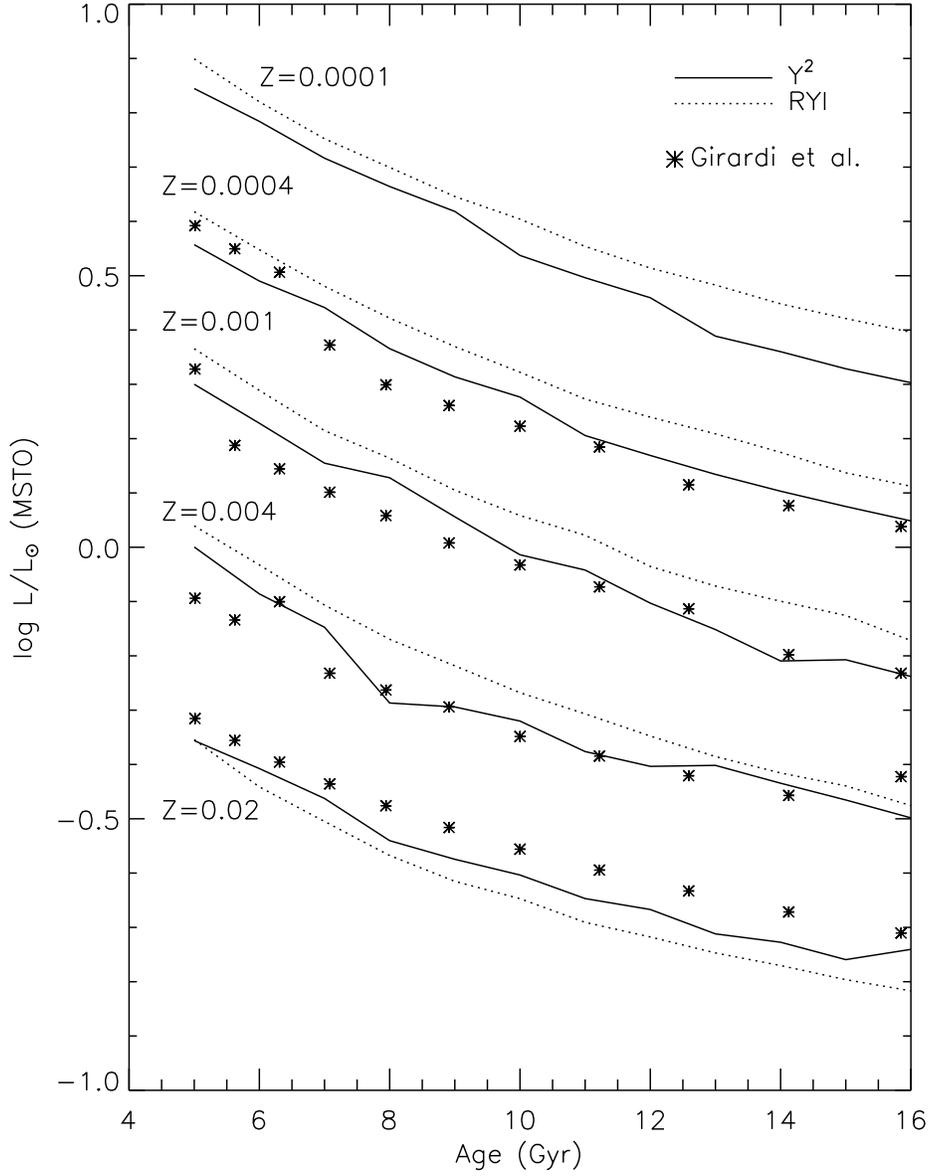}
\caption{The use of the updated stellar models alone leads to an age
reduction for the Galactic globular clusters by approximately 15\%.
A given observed MSTO brightness indicates a much smaller age when
the $Y^2$ isochrones are used than when the RYI are used.
The Girardi et al's (2000) isochrones are in good agreement with ours.
\label{fig14}}
\end{figure}

\begin{figure}
\epsscale{1}
\plotone{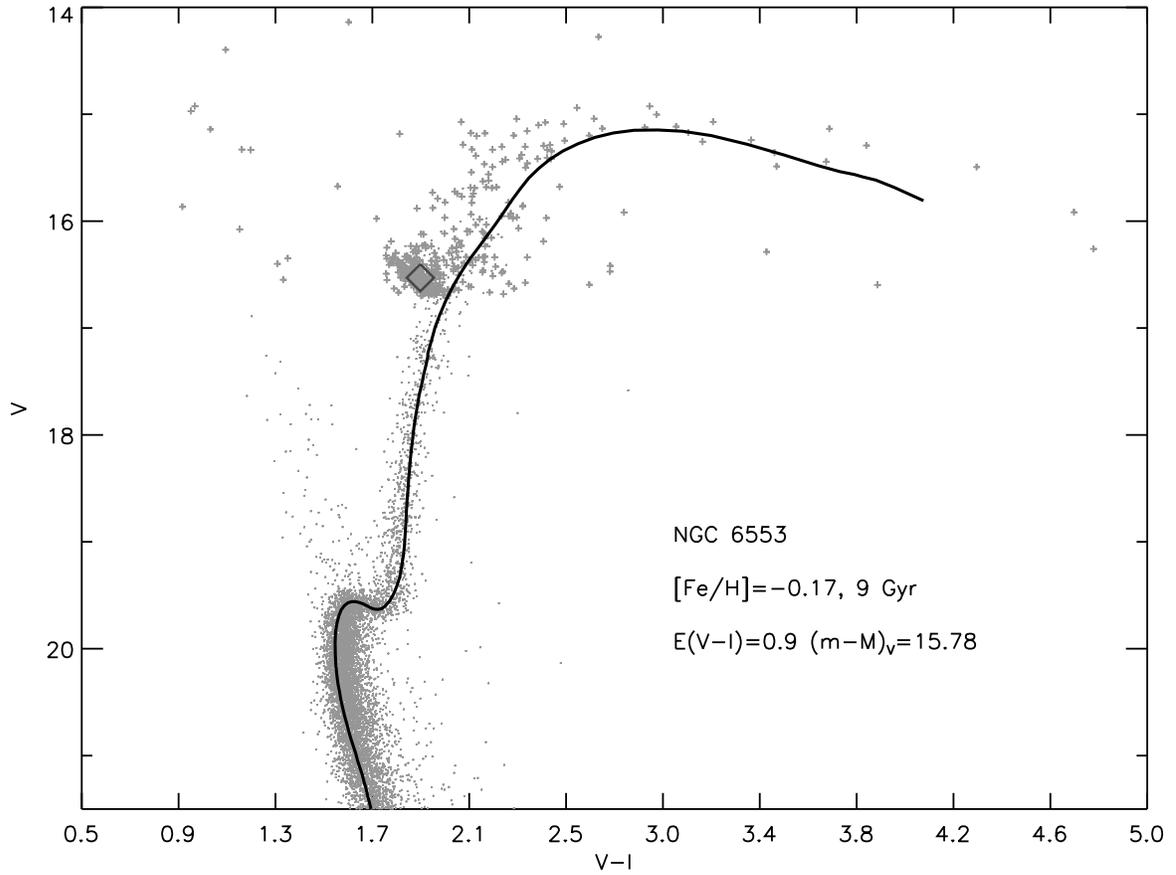}
\caption{
A test to the HRD of the metal-rich cluster NGC\,6553.
The new isochrone (based on the LCB color table) matches the overall
HRD well at 9\,Gyr. The data are from Zoccali et al. (2001; small dots) and
from Sagar et al. (1999; crosses). The open diamond is from our
synthetic HB models. 
\label{fig15}}
\end{figure}

\begin{figure}
\plotone{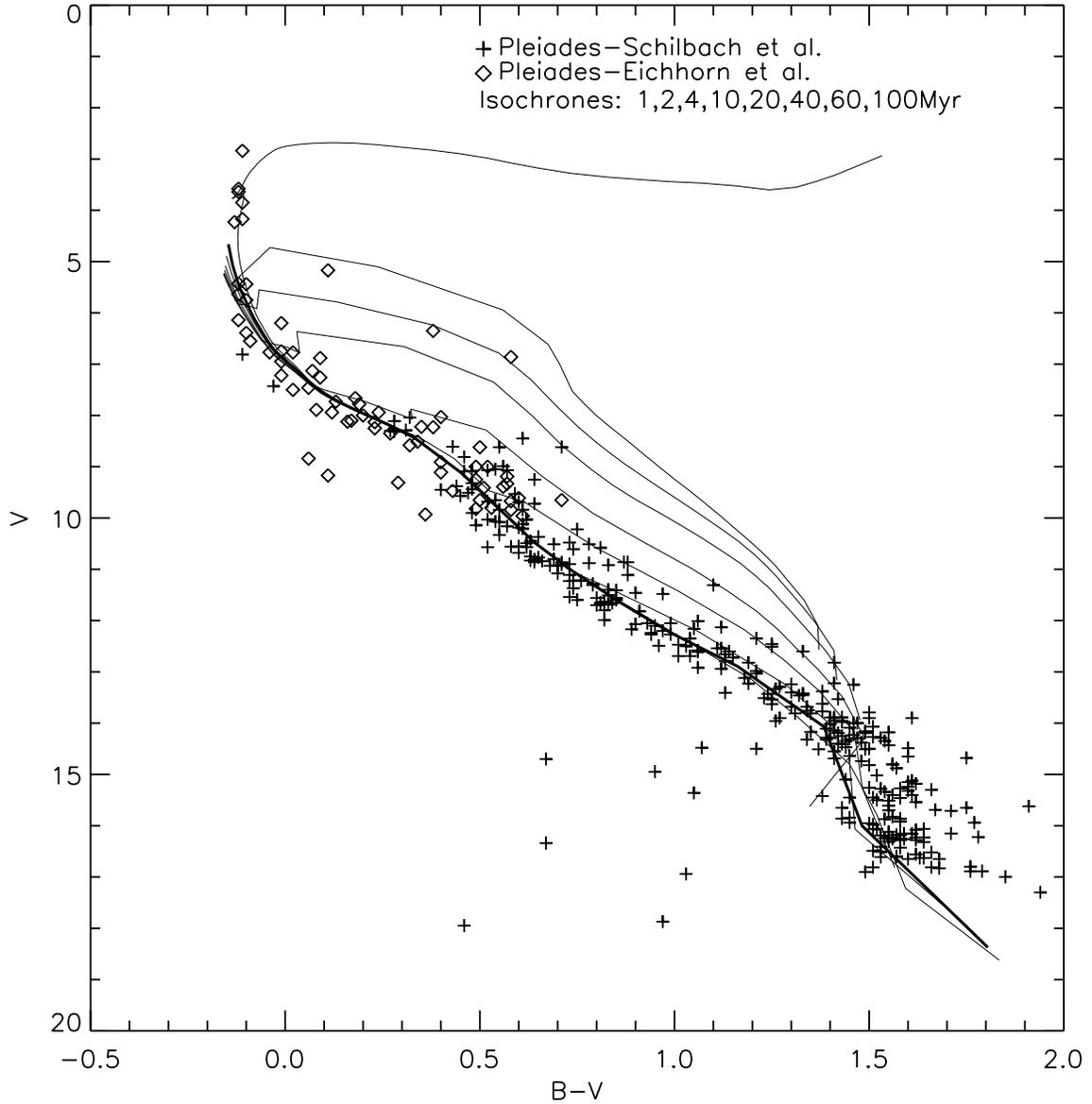}
\caption{A test to the HRD of Pleiades. The data are from Schilbach et al.
(1998, crosses) and from Eichhorn et al. (1970, diamonds). The Eichhorn
et al. data are truncated at $V=10$ mag, because they are overlapped
by Schilbach et al's more recent data. Isochrones are for $Z=0.02$ and for
ages of 1, 2, 4, 10, 20, 40, 60 (thick line), 100 (top) Myr from right to left.
Adopted parameters are (m-M)=5.6, E(B-V)=0.04, and $R_{V}=3.1$.
\label{fig16}}
\end{figure}

\begin{figure}
\plotone{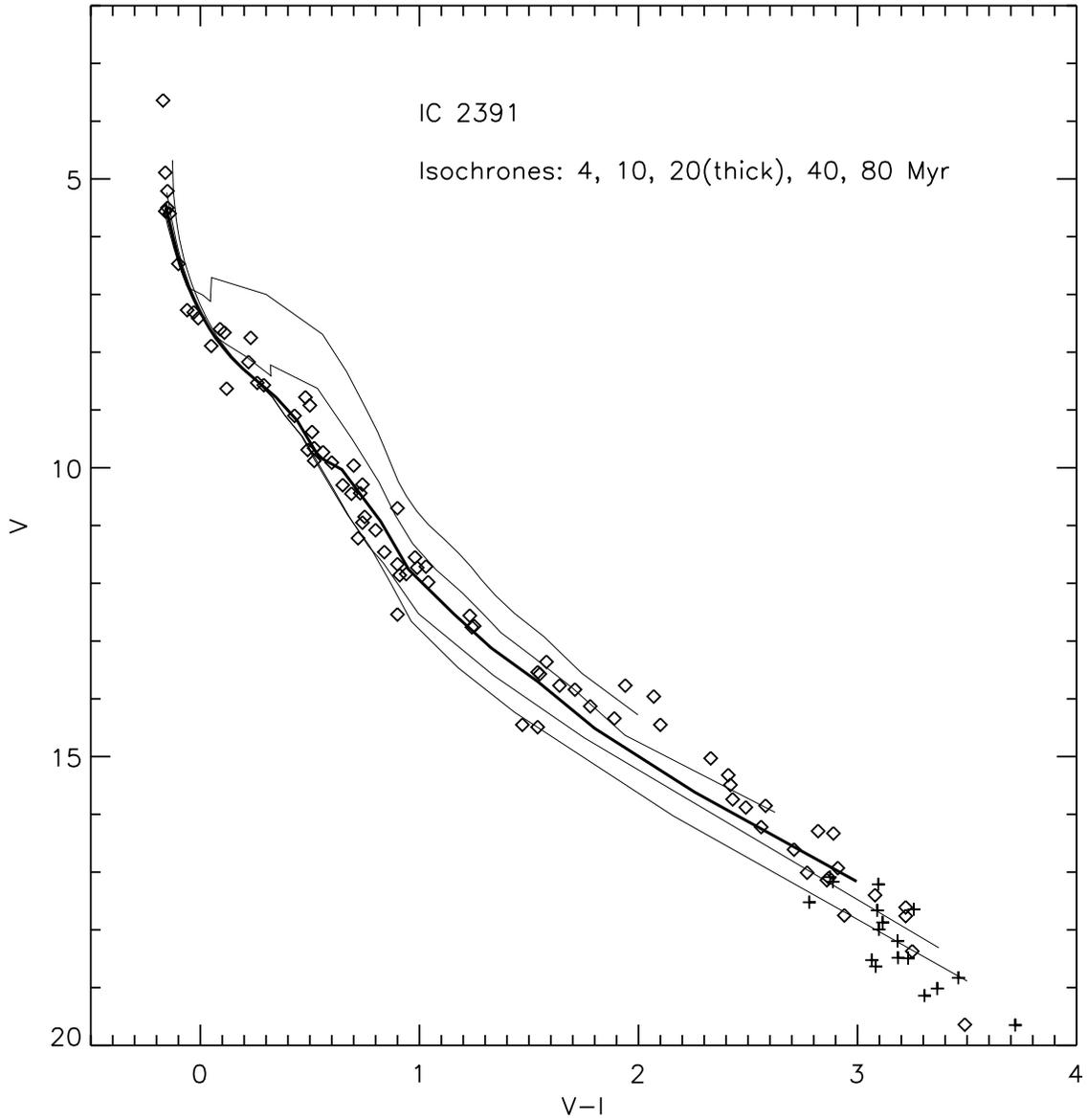}
\caption{A test to the HRD of IC\,2391. This young open cluster is
matched by the 20Myr solar-composition isochrone. The data are from
Patten \& Pavlovsky (1999) and have been kindly provided by Patten.
Adopted parameters are (m-M)=6.05, E(B-V)=0.01, and $R_{V}=3.1$.
\label{fig17}}
\end{figure}

\begin{figure}
\plotone{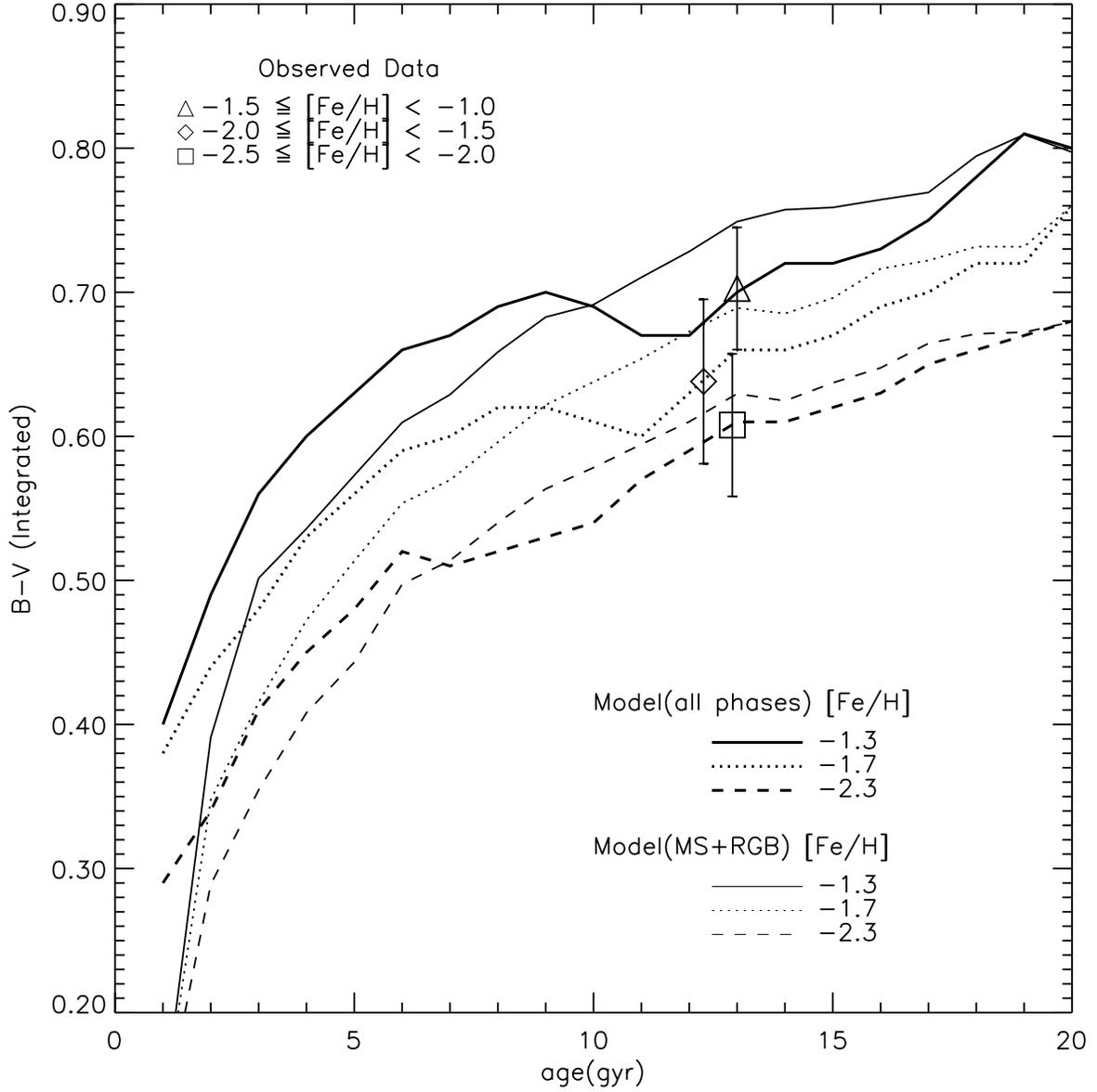}
\caption{The evolution in integrated color of metal-poor populations.
Compared are the observed integrated color of Galactic globular clusters
with $E(B-V)<1.0$. The data are from Harris (1996). Thine lines are
the models that contain only MS and RGB stars. Thick lines are the models
that contain all evolutionary phases. The age estimates indicated by
their integrated colors are in good agreement with isochrone-derived ages.
\label{fig18}}
\end{figure}

\end{document}